\documentclass[aps,amsmath,amssymb,nofootinbib,
reprint,longbibliography,superscriptaddress,
preprintnumbers,prd]{revtex4-2}
\pdfoutput=1
\usepackage{graphicx}
\usepackage{subcaption}
\usepackage{lipsum}
\usepackage{booktabs}
\usepackage{tabularx}
\usepackage{tikz}
\usepackage{pgfplots}
\usetikzlibrary{positioning,shapes,trees,arrows,graphs,decorations,calc}
\usepackage{url}

\usepackage{amssymb}
\usepackage{amsmath}
\usepackage{hyperref}
\usepackage{xcolor}

\usepackage{placeins} 
\usepackage[normalem]{ulem}
\usepackage{xspace}

\newcommand{\eq}[1]{eq.~\eqref{eq:#1}}
\newcommand{\Eq}[1]{Eq.~\eqref{eq:#1}}
\newcommand{\eqs}[2]{eqs.~\eqref{eq:#1} and \eqref{eq:#2}}
\renewcommand{\sec}[1]{sec.~\ref{sec:#1}}

\newcommand{\fig}[1]{Fig.~\ref{fig:#1}}

\newcommand{\app}[1]{app.~\ref{app:#1}}

\newcommand{\refcite}[1]{ref.~\cite{#1}}

\newcommand{\ord}[1]{\mathcal{O}(#1)}

\newcommand{\df}{\mathrm{d}}
\newcommand{\img}{\mathrm{i}}

\newcommand{\as}{\alpha_{s}}

\newcommand{\Tau}{\mathcal{T}}

\newcommand{\nn}{\nonumber}

\newcommand{\FO}{\mathrm{FO}}

\newcommand{\de}{\mathrm{d}}
\newcommand{\cusp}{\mathrm{cusp}}

\newcommand{\cL}{\mathcal{L}}

\newcommand{\geneva}{\textsc{Geneva}\xspace}

\newcommand{\openloopsTwo}{\textsc{OpenLoops2}\xspace}

\newcounter{bla}

\newcommand{\rescaletwoplots}{0.49\textwidth}
\newcommand{\hspacebetweentwoplots}{0em}

\newcommand{\spaceabovefigurecaption}{-5ex}

\makeatletter
\g@addto@macro\bfseries{\boldmath}
\makeatother

\DeclareMathAlphabet\mathbfcal{OMS}{cmsy}{b}{n}

\begin{document}


\title{N$^3$LL resummation of one-jettiness for $Z$-boson
  plus jet\\ production at hadron colliders}

\author{Simone Alioli}
\affiliation{Universit\`a degli Studi di Milano-Bicocca \& INFN
  Sezione di Milano-Bicocca, Piazza della Scienza 3, Milano 20126,
  Italy}

\author{Guido Bell}
\affiliation{Theoretische Physik 1, Center for Particle Physics Siegen, Universit\"at Siegen, Germany}

\author{Georgios Billis}
\affiliation{Universit\`a degli Studi di Milano-Bicocca \& INFN
  Sezione di Milano-Bicocca, Piazza della Scienza 3, Milano 20126,
  Italy}

\author{Alessandro Broggio}
\affiliation{Faculty of Physics, University of Vienna, Boltzmanngasse 5, A-1090 Wien, Austria}

\author{Bahman Dehnadi}
\affiliation{Deutsches Elektronen-Synchrotron DESY, Notkestr. 85, 22607 Hamburg, Germany}

\author{Matthew~A.~Lim}
\affiliation{Department of Physics and Astronomy, University of Sussex, Sussex House, Brighton, BN1 9RH, UK}

\author{Giulia Marinelli}
\affiliation{Universit\`a degli Studi di Milano-Bicocca \& INFN
  Sezione di Milano-Bicocca, Piazza della Scienza 3, Milano 20126,
  Italy}
\affiliation{Deutsches Elektronen-Synchrotron DESY, Notkestr. 85, 22607 Hamburg, Germany}

\author{Riccardo Nagar}
\affiliation{Universit\`a degli Studi di Milano-Bicocca \& INFN
  Sezione di Milano-Bicocca, Piazza della Scienza 3, Milano 20126,
  Italy}

\author{Davide Napoletano}
\affiliation{Universit\`a degli Studi di Milano-Bicocca \& INFN
  Sezione di Milano-Bicocca, Piazza della Scienza 3, Milano 20126,
  Italy}

\author{Rudi Rahn}
\affiliation{Department of Physics and Astronomy, University of Manchester,
  Manchester, M13 9PL, UK}

\preprint{SI-HEP-2023-31}
\preprint{P3H-23-098}
\preprint{DESY-23-206}
\preprint{UWThPh-2023-29}
\date{\today}

\begin{abstract}
We present the resummation of one-jettiness for the colour-singlet plus jet production process $p p \to  ( \gamma^*/Z \to \ell^+ \ell^-) + {\text{jet}}$ at hadron colliders up to the fourth logarithmic order (N$^3$LL). This is the first resummation at this order for processes involving three coloured partons at the Born level. We match our resummation formula to the corresponding fixed-order predictions, extending the validity of our results to regions of the phase space where further hard emissions are present. This result paves the way for the construction of next-to-next-to-leading order simulations for colour-singlet plus jet production matched to parton showers in the \geneva framework.
\end{abstract}

\maketitle

\section{Introduction}

The study of the production of a colour singlet system  at large recoil is of crucial importance for the physics programme at the Large Hadron Collider. In particular, theoretical predictions for $\gamma^*/Z+$jet production are needed at higher precision to match the accuracy reached by experimental measurements of the $Z$ boson transverse momentum ($q_T$) spectrum. Combining next-to-next-to-leading order (NNLO) predictions for $\gamma^*/Z+$jet~\cite{Gehrmann-DeRidder:2015wbt,Gehrmann-DeRidder:2016cdi,Gehrmann-DeRidder:2016jns,Gehrmann-DeRidder:2017mvr,Gehrmann-DeRidder:2019avi,Gauld:2020deh} with $q_T$ resummation~\cite{Bizon:2018foh,Bizon:2019zgf,Re:2021con,Camarda:2021ict,Chen:2022cgv,Neumann:2022lft,Camarda:2023dqn, Billis:2023xxx} provides an accurate description of this distribution over the whole kinematic range and can be used to extract $\alpha_s$~\cite{ATLAS:2023lhg} and as a background for new physics searches.

The one-jettiness variable is a suitable event shape for colour singlet ($L$) + jet production which does not suffer from superleading or nonglobal logarithms. It is a specific case of $N$-jettiness~\cite{Stewart:2010pd}, and has been used to perform slicing calculations at NNLO~\cite{Gaunt:2015pea,Boughezal:2015dva,Boughezal:2015ded,Boughezal:2015aha,Campbell:2019gmd}. Resummation of the jettiness has been performed for various $N$~\cite{Alioli:2015toa,Alioli:2020fzf,Jouttenus:2013hs,Alioli:2021ggd}, and this was exploited to match NNLO calculations to parton shower algorithms for colour singlet production in \geneva~\cite{Alioli:2015toa,Alioli:2019qzz,Alioli:2020fzf,Alioli:2020qrd,Alioli:2021egp,Alioli:2022dkj,Alioli:2023har}. In this work, we resum the one-jettiness up to N$^3$LL accuracy, providing state-of-the-art predictions for this variable, which was only previously known up to NNLL~\cite{Jouttenus:2013hs}.
In order to obtain this accurate result, we rely on higher-order perturbative ingredients which have only become available in the last few years. In particular, the structure of the hard anomalous dimensions that is relevant for N$^3$LL resummation was derived in \refcite{Becher:2019avh} together with the direct evaluation of the four-loop cusp anomalous dimension in \refcite{Henn:2019swt, vonManteuffel:2020vjv}. N$^3$LL resummation also requires the knowledge of two-loop soft boundary terms which were first evaluated in \refcite{Boughezal:2015eha,Campbell:2017hsw} and recomputed for this paper with a refined treatment of the small and large angle regions \cite{BDMR}.

We define the one-jettiness resolution variable as \cite{Stewart:2010pd}
\begin{align}\label{eq:tau1def}
\Tau_1 = \sum_k \min \bigg\{ \frac{2 q_a \cdot p_k}{Q_a}, \frac{2 q_b \cdot p_k}{Q_b}, \frac{2 q_J \cdot p_k}{Q_J} \bigg\}\,,
\end{align}
with $q_{a, b} = x_{a, b} E_{\mathrm{cm}}\, n_{a, b}/2 = E_{a,b}\, n_{a, b}$ and \mbox{$q_J = E_J\, n_J$}, where $E_J$ is the jet energy. The beam directions are  $n_{a, b}=(1,0,0,\pm 1)$ while the massless jet direction is $n_J=(1,\vec{n}_J)$.
In \eq{tau1def} the sum runs over the four-momenta $p_k$ of all partons which are part of the hadronic final state.
We use a geometric measure where \mbox{$Q_i=2 \rho_i E_i$}  with $i = a,b,J$ is proportional to the energy of the beam or jet momenta.
This particular choice is preferable because it is independent of the total jet energy, but makes the one-jettiness definition frame dependent. Results in frames  that differ by a longitudinal boost can
be obtained by making different  choices  for $\rho_i$.
In this work we show results for $\Tau_1$ in the laboratory frame (LAB)
and in the  frame where the colour singlet system has zero rapidity (CS).
The LAB frame definition is obtained by setting  $\rho_i=1$ and evaluating the
jet energy and the directions of the partonic momenta in the laboratory.
In order to obtain the  CS frame definition we instead set
\begin{align}
\rho_a = e^{\hat{Y}_L},\, \rho_b = e^{-\hat{Y}_L},\, \nn
 \rho_J = (e^{-\hat{Y}_L} \hat{q}_J^+ + e^{\hat{Y}_L} \hat{q}_J^-)/(2 \hat{E}_J) ,
\end{align}
where $\hat{Y}_{L}$ is the rapidity of $L$ in the laboratory. The quantities $\hat{q}_J^\pm= \hat{q}_J^0 \mp \hat{q}_J^3$ and $\hat{E}_J$ are the lightcone components and energy of the reconstructed massless jet four-momentum $\hat{q}_J$ in the laboratory frame respectively.
In this way the longitudinal boost between the two frames is absorbed by a redefinition of the $\rho_i$.

The manuscript is organised as follows.
In \sec{theory} we introduce the  factorisation
formula, detailing its ingredients and their renormalisation group
(RG) evolution. We present
a final resummed formula valid up to N$^3$LL accuracy and we match it with the
appropriate fixed-order calculation in order to extend the description of
the one-jettines spectrum  also in regions where more than one hard jet is present.
In \sec{implementation} we  discuss the details of the implementation and
present our results for the one-jettiness distribution.  We also study the
nonsingular contribution in different frames and provide
predictions matched to the appropriate fixed-order (FO) distributions.
We finally draw our conclusions in \sec{conclusions}. 
Further details about the derivation of the resummed results are described in
the appendices.
  
\section{Factorisation and Resummation}
\label{sec:theory}

A general factorisation formula for the $N$-jettiness distribution was derived in \refcite{Stewart:2009yx,Stewart:2010tn}. For the case of one-jettiness in hadronic collisions it reads
\begin{align}\label{eq:fact}
    \frac{\de\sigma}{\de\Phi_1\de\Tau_1}\! =\! & \sum_{\kappa}H_{\kappa}(\Phi_1, \mu) \int\, \de t_a\, \de t_b\, \de s_J
     \\  & \ \times \  B_{\kappa_a}( t_a, x_a,  \mu)\ B_{\kappa_b}(t_b, x_b,  \mu)\  J_{\kappa_J}(s_J, \mu)\nn
        \\  & \ \times \ S_{\kappa}\left(n_a\cdot n_J,\Tau_1-\frac{t_a}{Q_a}-\frac{t_b}{Q_b}-\frac{s_J}{Q_J}, \mu\right)\nn
\,,
\end{align}
where $x_{a,b}= (Q_{LJ} / E_{\mathrm{cm}}) \exp\{\pm Y_{LJ}\}$ and $Q_{LJ}$ is
the invariant mass of the colour-singlet plus jet system ($LJ$). The index set
$\kappa \equiv \{\kappa_a, \kappa_b, \kappa_J\}$ runs over all allowed
partonic channels and $\kappa_a$, $\kappa_b$, $\kappa_J$  denote the
individual parton types.~$\Phi_1$ is the phase space for the $LJ$ system and
$n_a\cdot n_J = (1-\cos \theta_{aJ})$ measures the angle between the jet and
the rightward beam direction in the laboratory frame.
In general, for $L$+jet production  all permitted partonic channels contribute, i.e. $\kappa_a \,\kappa_b\, \kappa_J \in \{q\bar{q}g, qgq, ggg, \ldots\}$, where we have indicated all the crossing and charge-conjugated processes within the dots.
For the $p p \to  ( \gamma^*/Z \to \ell^+ \ell^-) + {\text{jet}}+X$ case we consider in this work, the $q\bar{q}g$ and $qgq$ channels (plus their crossing and charge-conjugated ones) appear at Born level.
The $ggg$ channel instead begins to contribute only at $\ord{\as^3}$.

In \eq{fact} the hard functions $H_\kappa$ are defined as the square of the
Wilson coefficients of the effective theory operators defined in
Soft-Collinear Effective Theory (SCET). They can be obtained from the  UV- and IR-finite relevant amplitudes in  full QCD. The beam $B_{\kappa_{a/b}}$ and the jet $J_{\kappa_J}$ functions describe collinear emissions along the beam and jet directions respectively.
The functions $S_\kappa$ describe isotropic soft emissions from soft Wilson
lines and depend on the angle between the beam and jet directions.

When the hard, soft, beam and jet functions are evaluated at a common scale $\mu$, large logarithms of the ratios of disparate scales may arise, which spoil the convergence of fixed-order perturbation theory.
The resummation of such logarithms is achieved through RG evolution in the SCET framework. All the functions appearing in the factorisation formula are evolved from their characteristic energy scales ($\mu_X,\, X= H,S,B,J$) to the common scale $\mu$ by separately solving their associated RG evolution equations.
The accuracy of the resummed predictions is systematically improvable by
including higher-order terms in the fixed-order expansions of the hard, soft,
beam and jet functions as well as in their corresponding anomalous
dimensions. To achieve N$^3$LL accuracy one needs the boundary conditions of
the hard, soft, beam and jet functions up to two loops. The coefficients of
the scale-dependent and kinematic-dependent logarithmic terms in the anomalous
dimension and the QCD beta function need to be evaluated up to four
loops. Finally, nonlogarithmic noncusp terms in the anomalous dimension need to be evaluated up to three loops.

In the rest of this section we will present the functions appearing in the
factorisation formula~\eqref{eq:fact} and their
evolution separately and derive the final resummed formula in \sec{resummed_formula}.

\subsection{Hard functions for $p p \to  ( \gamma^*/Z \to \ell^+ \ell^-) + {\text{jet}}$} 

The hard function for the channel $\kappa$
satisfies the following RG equation (RGE)
\begin{align}
\frac{\de}{\de \log \mu} H_\kappa(\Phi_1,\mu) &= \Gamma_H^\kappa(\mu) \,
                                         H_\kappa(\Phi_1,\mu) \,,
\end{align}
with $\Gamma^\kappa_H(\mu) = 2 \mathrm{Re} \,\{ \Gamma^\kappa_C(\mu)\}$.
Here we have already exploited the fact that for the colour-singlet plus jet production process, the colour structure is trivial, i.e. the anomalous dimensions of the Wilson coefficient $\Gamma^\kappa_C(\mu)$ (or equivalently the anomalous dimension of the hard function  $\Gamma^\kappa_H(\mu)$) is diagonal in colour space, as we show below.
For ease of notation we use in this section the abbreviations $a= \kappa_a$, $b= \kappa_b$ and $c= \kappa_J$.
Writing the  anomalous dimension $\Gamma^\kappa_C(\mu)$ in full generality as
a matrix in colour space and using its explicit expression up to N$^3$LL given
in \refcite{Becher:2019avh}, we find
\begin{widetext}
\vspace{-0.80cm}
  \begin{align} \label{eq:hardfuncn3llandim}
	\boldsymbol{\Gamma}_C^\kappa(\mu) 
	&=  \Gamma_C^\kappa(\mu) \	\boldsymbol{1} \ = 
	\bigg\{ \frac{\Gamma_\cusp(\as)}{2} \bigg[ \big(C_c - C_a - C_b\big) \ln\frac{\mu^2}{(-s_{ab}-\mathrm{i} 0)} + \textrm{cyclic permutations} \bigg]
	\nn \\ & \quad
	+ \gamma^a_C(\as) + \gamma_C^b(\as) + \gamma_C^c(\as) + {{\frac{C^2_A}{8}}}f(\as)\big(C_a + C_b + C_c\big) \bigg\} \boldsymbol{1}
	\nn \\ & \quad
	+ \sum_{(i,j)} \bigg[ -f(\as) \mathbfcal{T}_{iijj}
	+ \sum_{R=F,A} g^R(\as) \big(3\mathbfcal{D}^R_{iijj} + 4\mathbfcal{D}^R_{iiij}\big) \ln\frac{\mu^2}{(-s_{ij} - \mathrm{i} 0)} \bigg] + \ord{\as^5}
                 \,,
\end{align}
\end{widetext}
where the sums run over all the external hard parton pairs with $i \neq j$ and  $C_{i}$ is the quadratic Casimir invariant for the parton $i$ in the colour representation $R_i$. The symbol `$\boldsymbol{1}$' denotes the identity element in colour space.
The cusp $\Gamma_\cusp(\as)$ and noncusp $\gamma^i_C(\as)$ anomalous dimensions are given in
App.~A of \refcite{Becher:2019avh}
 for both quark and gluon cases.\footnote{In the notation of \refcite{Becher:2019avh} they read $\Gamma_\cusp(\as)\equiv\gamma_\cusp(\as)$ and $\gamma^{i}_{C}(\as)\equiv \gamma^{i}(\as)$.}
We have $\Gamma_\cusp(\as) =  \sum_{n=0}^\infty \left(\frac{\as} {4 \pi}\right)^n \Gamma_n $, with $\Gamma_n$ the universal cusp anomalous dimension coefficients.
The symmetrised colour structures that appear in \eq{hardfuncn3llandim} are defined as
\begin{align}
\mathbfcal{T}_{ijkl}
&= f^{ade}f^{bce} (\boldsymbol{T}^a_i \boldsymbol{T}^b_j \boldsymbol{T}^c_k \boldsymbol{T}^d_l)_+
\,, \nonumber\\
\mathbfcal{D}^R_{ijkl}
&= d^{abcd}_R \boldsymbol{T}^a_i \boldsymbol{T}^b_j \boldsymbol{T}^c_k \boldsymbol{T}^d_l\, ,
\end{align}
where $\big(\boldsymbol{T}^{a_1}_{i_1}\ldots \boldsymbol{T}^{a_n}_{i_n}\big)_+\equiv\frac{1}{n!}\sum_{\pi} \boldsymbol{T}^{a_{\pi(a)}}_{i_{\pi(1)}} \ldots \boldsymbol{T}^{a_{\pi(n)}}_{i_{\pi(n)}} $ denotes the normalised sum of all possible permutations $\pi$ of the $n$ colour operators and
\begin{align}
  d^{a_1\ldots a_n}_R &= \mathrm{Tr}_R(\boldsymbol{T}^{a_1} \ldots \boldsymbol{T}^{a_n})_+\nn\\&= \frac{1}{n !}\sum_{\pi} \mathrm{Tr}\big(\boldsymbol{T}^{a_{\pi(1)}}_R\ldots \boldsymbol{T}^{a_{\pi(n)}}_R\big)\,.
  \end{align}

The functions $f(\as)$ and $g^{R}(\as)$ ($R=F$ for the fundamental and $R=A$ for the adjoint
representation) start at $\ord{\as^3}$ and $\ord{\as^4}$ respectively. The explicit expressions can be derived from  \refcite{Becher:2019avh, Henn:2019swt, vonManteuffel:2020vjv}; we report them below for completeness
\begin{align}
   f(\as)  = &\ 16\, \big(\, \zeta_5 + 2 \zeta_2 \zeta_3\, \big)\, \left( \frac{\as}{4 \pi}\right)^3  + \ord{\as^4} \nn \\
  g^F(\as)  = &\ T_F\, n_f \left(\frac{128 \pi^2}{3} -\frac{256 \zeta_3}{3} - \frac{1280 \zeta_5}{3} \right) \left( \frac{\as}{4 \pi}\right)^4 \nn \\ &\ + \ord{\as^5} \\
    g^A(\as)  = &\ \left( -64\, \zeta_2 - \frac{3968}{35}\, \zeta_2^3 + \frac{64}{3}\, \zeta_3 - 192\, \zeta_3^2 \right. \nn \\ & \left.\qquad  + \frac{1760}{3} \zeta_5  \right) \left( \frac{\as}{4 \pi}\right)^4  + \ord{\as^5} \,.\nn
\end{align}
The terms proportional to these functions start contributing only at N$^3$LL accuracy. In particular, similar  to the  $\Gamma_\cusp(\as)$ case,  $g^R(\as)$  needs to be known one order higher than $f(\as)$ since it multiplies a scale logarithm.

It is possible to show using colour conservation relations ($\sum_{i=a,b,c}
\boldsymbol{T}_i |\mathcal{M} \rangle = 0$) and the symmetry properties of
$d_R^{abcd}$ that a symmetric combination of the term proportional to
$g^R(\as)$ can be rewritten in terms of quartic Casimirs
\begin{align}
  C_4(R_i,R) = \frac{d_{R_i}^{abcd}d_{R}^{abcd}}{N_{R_i}} \equiv D_{iR} \, ,
  \end{align}
  associated to the external legs, where $N_{R_i}$ is the dimension of the colour representation $R_i$ (i.e. $N_F= N_c$  and $N_A=N_c^2-1$ for the fundamental and adjoint representations of $SU(N_c)$ respectively).
  The explicit form of the $ D_{iR} $ is
  \begin{align}
    D_{FF} & = \frac{(N_c^4 - 6 N_c^2 + 18) (N^2_c-1)}{96 N_c^3}\,, \nn \\
    D_{FA} & = \frac{(N_c^2 + 6) (N^2_c-1)}{48}\,, \\
    D_{AF} & = \frac{N_c (N_c^2 + 6)}{ 48}\,, \nn \\
    D_{AA} & = \frac{N_c^2 (N_c^2 + 36)}{24}\,. \nn 
    \end{align}
Similar relations can also be found by
exploiting consistency relations among anomalous dimensions. Explicitly, when
acting on the colour states we find
\begin{align}
  3\big(\mathbfcal{D}^R_{iijj}&+\mathbfcal{D}^R_{jjii}) +
                                4\big(\mathbfcal{D}^R_{iiij}+\mathbfcal{D}^R_{jjji}\big)
                                \nn\\&= (D_{kR}-D_{iR}-D_{jR})\ \mathbf{1}\,,
\end{align}
where $i \neq j \neq k$.
These relations have a similar structure to the quadratic Casimir case, where
for three coloured partons  one finds for example identities of the type 
\mbox{$\boldsymbol{T}_a\cdot
\boldsymbol{T}_b=[\boldsymbol{T}_c^2-\boldsymbol{T}_a^2-\boldsymbol{T}_b^2]/2$}. The
only relevant difference is the appearance of the index $R$ which labels the
fundamental and adjoint representations. This is due to the presence of different partons in the internal loops.
We have verified that these relations hold by directly evaluating the action
of the colour insertion operators on the possible colour states in the
colour-space formalism. We have further checked these relations using the
\texttt{ColorMath} package \cite{Sjodahl:2012nk}.

By employing these expressions, the logarithmic term of the hard anomalous dimension in \eq{hardfuncn3llandim} can be further simplified and rewritten in terms of quartic Casimirs.
In order to do so we define 
\begin{align}
  \bar{c}^\kappa &=  c^\kappa_s + c^\kappa_u + c^\kappa_t =  - (C_{a}+C_{b} +C_{c})/2 
\,, \label{eq:gammacfull} \\
\bar{c}_L^\kappa &=  c^\kappa_s L_s + c^\kappa_u L_u + c^\kappa_t L_t 
                   \,, \label{eq:barc_L_def} \end{align}                
                 with
\begin{align}
c^\kappa_s = \boldsymbol{T}_a \cdot \boldsymbol{T}_b
\,,\quad c^\kappa_u = \boldsymbol{T}_b \cdot \boldsymbol{T}_c
\,,\quad 
c^\kappa_t = \boldsymbol{T}_a \cdot \boldsymbol{T}_c
\,.\end{align}
We also introduce an arbitrary hard scale $Q$ to separate the cusp and noncusp terms and  use the abbreviations 
\begin{align}
L_s &= \ln\frac{-s_{ab} - \img 0}{Q^2}  = \ln\frac{s_{ab}}{Q^2} - \img \pi
\,,\nn \\
L_u &= \ln\frac{s_{bc}}{Q^2}\,, \qquad L_t  = \ln\frac{s_{ac}}{Q^2}\,.\nn
\end{align}
By  analogy to the quadratic case, we also define the sum of the quartic Casimirs of the external coloured legs as
\begin{align}
    \bar{c}^{\kappa, \, R}_4 = D_{aR} + D_{bR} +D_{cR} \, .
\end{align}
For the quartic Casimir terms the kinematic dependence is encoded by 
\begin{align}
\bar{c}_{4,\, L}^{\, \kappa,\, R} \equiv c^{\kappa,\, R}_{4,\, s}\, L_s + c^{\kappa,\, R}_{4,\, u} \, L_u + c^{\kappa,\, R}_{4,\, t} \, L_t\,,
\end{align}
where
\begin{align}
c^{\kappa,\, R}_{4,\, s} =  D_{aR} + D_{bR} -D_{cR} \,, \nn \\
c^{\kappa,\, R}_{4,\, t} =  D_{aR} + D_{cR} -D_{bR}\,, \\
c^{\kappa,\, R}_{4,\, u} =  D_{bR} + D_{cR} -D_{aR}\,. \nn
\end{align}
Using all the above definitions the anomalous dimension of the Wilson coefficient for each channel $\kappa$  can be written in a fully diagonal form in colour space as
\begin{align}
\Gamma^\kappa_C(\mu) = & \left[   -\bar{c}^\kappa \Gamma_\cusp(\as) +
                       \sum_{R=F,A}\,\bar{c}^{\kappa, \, R}_4 \, g^R(\alpha_s)
                       \right] \ln\frac{Q^2}{\mu^2} \, \nn \\
                       & + \sum_{i=a,b,c}\gamma_C^i(\as) + f(\as) \, c^\kappa_{f}  -\bar{c}_L^\kappa \Gamma_\cusp(\as)  \, \nn \\
                       & + \sum_{R=F,A}\, g^R(\alpha_s) \,\bar{c}_{4,\, L}^{\, \kappa,\, R}    \,,
\end{align}
where  the  last missing ingredient appearing in the noncusp anomalous dimensions is
\begin{align}
c^\kappa_{f} =  -\biggl[ {\frac{C^2_A}{4}} \, \bar{c}^{\kappa}
    + \sum_{i\neq j} \frac{\langle \mathcal{M} \vert \mathbfcal{T}_{iijj} \vert  \mathcal{M} \rangle}{\langle \mathcal{M} \vert \mathcal{M} \rangle} \biggr]\,.
\label{eq:gammaf}
\end{align}
This again requires an explicit evaluation of the action of the colour insertion operators on the possible colour states. 
We remind the reader that for three coloured partons the result of the colour
insertion operators must be diagonal and proportional to the identity by
Schur's lemma. Therefore, we consider their action on the amplitude in colour
space $\lvert\mathcal{M} \rangle$ for each partonic channel $\kappa$. The
colour amplitude  $\lvert\mathcal{M} \rangle$ is the same for all quark channels, $
\lvert \mathcal{M} \rangle
= t^a_{ji} \lvert i \, j \, a \rangle
$
where the $t^a_{ji}$ are the Gell-Mann matrices and the quantum numbers  $i \, (j)$ denote the colour of the quark (antiquark) and $a$ that of the gluon respectively.
We proceed by calculating separately for each channel the action of the colour operators as a function of the number of colours $N_c$.
For $\kappa = q \bar{q} g$  we find
\begin{align}
  \sum_{(i,j)}
\frac{\langle \mathcal{M} \lvert \mathbfcal{T}_{iijj} \rvert \mathcal{M}\rangle}{\langle \mathcal{M} \vert \mathcal{M}\rangle}
 &
= \frac{1}{\langle \mathcal{M} \vert \mathcal{M}\rangle}
\Bigl( 2 \langle \mathbfcal{T}_{qq\bar{q}\bar{q}} \rangle
+ 4 \langle \mathbfcal{T}_{qqgg} \rangle \Bigr)
\end{align}
where we used the abbreviation $\langle \mathbfcal{T}_{ijkl} \rangle \equiv
\langle \mathcal{M} \lvert \mathbfcal{T}_{ijkl} \rvert \mathcal{M} \rangle$ and
the relations
\begin{align}
\langle \mathbfcal{T}_{qq\bar{q}\bar{q}} \rangle
&= \langle \mathbfcal{T}_{\bar{q}\bar{q}qq} \rangle = \frac{3}{16} C_F N_c^2
  \,, \\
\langle \mathbfcal{T}_{xxgg} \rangle &= \langle \mathbfcal{T}_{ggxx} \rangle = \frac{1}{16} C_F N_c^2 (N_c^2 + 4) \,, \qquad x = q, \bar{q} \nn
\,. \end{align}
The normalisation factor corresponds to the colour factor of the
Born amplitude $\langle \mathcal{M} \vert \mathcal{M}\rangle = C_F N_c$.

For the $\kappa = q g q$ channel it is crucial to properly take into account whether the quark is in the initial state or in the final state,
since it uniquely defines the action of the colour operators on the colour states. We do so by using the notation $q_i$ ($q_f$) for the initial (final) state quark. We find
\begin{align}
\sum_{(i,j)}
\frac{\langle \mathcal{M} \lvert \mathbfcal{T}_{iijj} \rvert \mathcal{M}\rangle}{\langle \mathcal{M} \vert \mathcal{M}\rangle}
&
= \frac{1}{\langle \mathcal{M} \vert \mathcal{M}\rangle}
\Bigl( 2 \langle \mathbfcal{T}_{q_i q_i q_f q_f} \rangle
+ 4 \langle \mathbfcal{T}_{q_i q_i gg} \rangle \Bigr)
\,,\end{align}
where  we used
\begin{align}
\langle \mathbfcal{T}_{q_i q_i q_f q_f} \rangle
&\equiv \langle \mathbfcal{T}_{q_f q_f q_i q_i} \rangle
= C_F N_c^2 \frac{3}{16}
\,, \\
\langle \mathbfcal{T}_{q_x q_x gg} \rangle
& \equiv \langle \mathbfcal{T}_{gg q_x q_x} \rangle
= C_F N_c^2 \frac{N_c^2 + 4}{16} \nn
\,, \qquad x = i, f
\,.\end{align}
Finally, the $\kappa = \bar{q} g \bar{q}$ can be obtained trivially  from the $\kappa = q g q$ results simply by applying charge conjugation and replacing the quark with an antiquark.
Some of these colour factors also appear in the calculation of the threshold three-loop soft function in \refcite{Liu:2020wmp}, for which we find complete agreement.

Everything is now in place to write the solution of the RGE for the hard Wilson coefficient. Indicating with $\mu_H$ its canonical scale, the evolution kernel for the hard function  $U^\kappa_H(\mu_H,\mu) = |  U^\kappa_C(\mu_H,\mu) |^2$ reads
\begin{widetext}
\begin{align}
U^\kappa_H(\mu_H,\mu) &= \exp \bigg[4\, \bar{c}^{\kappa} K_{\Gamma_{\mathrm{cusp}}}(\mu_H,\mu) - 4 \,\bigg(\sum_{R=F,A }\bar{c}^{\kappa,\, R}_4 K_{g^R} (\mu_H, \mu)\bigg)    - 2\bar{c}^{\kappa} \eta_{\Gamma_{\mathrm{cusp}}}(\mu_H,\mu) \ln \frac{Q^2}{\mu^2_H} \, \nn\\
& + 2 \bigg(\sum_{R=F,A }\bar{c}^{\kappa,\, R}_4 \eta_{g^R}(\mu_H,\mu) \bigg)  \ln \frac{Q^2}{\mu^2_H} - 2 {\rm Re}\left\{\bar{c}^{\kappa}_L\right\} \, \eta_{\Gamma_{\mathrm{cusp}}}(\mu_H,\mu) +   \sum_{R=F,A}\, 2 {\rm Re}\left\{\bar{c}_{4,\, L}^{\, \kappa,\, R}\right\}\, \eta_{g^R}(\mu_H,\mu) \, \nn\ \\
& + 2 \sum_{i=a,b,c} K_{\gamma^i_C}(\mu_H,\mu) + 2 c^\kappa_f \, K_{f}(\mu_H,\mu)\bigg] \, ,
\end{align}
\end{widetext}
~\\~\\~\\
where we have used the definitions
\begin{align}
K_{\Gamma_x}(\mu_H, \mu) &= \int^{\as(\mu)}_{\as(\mu_H)} \frac{\de\as}{\beta(\as)} \Gamma_x(\as)
    \int_{\as(\mu_H)}^{\as} \frac{\de\as^\prime}{\beta(\as^\prime)}
\,, \nn\\
\eta_{\Gamma_x}(\mu_H, \mu) &= \int_{\as(\mu_H)}^{\as(\mu)} \frac{\de \as}{\beta(\as)} \Gamma_x(\as)
\,, \\
K_{\gamma_x}(\mu_H, \mu) &= \int_{\as(\mu_H)}^{\as(\mu)} \frac{\de
                           \as}{\beta(\as)} \gamma_x(\as) \nn
\end{align}
and
\begin{align}
K_{g^R} (\mu_H, \mu) &\equiv  \int^{\as(\mu)}_{\as(\mu_H)}
                       \frac{\de\as}{\beta(\as)} g^R(\as)
                       \int_{\as(\mu_H)}^{\as}
                       \frac{\de\as^\prime}{\beta[\as^\prime]} \,, \nn \\
\eta_{g^R} (\mu_H, \mu) &\equiv \int_{\as(\mu_H)}^{\as(\mu)} \frac{\de \as}{\beta(\as)} g^R(\as)  \,,\\
K_{f}(\mu_H, \mu) & \equiv \int_{\as(\mu_H)}^{\as(\mu)} \frac{\de \as}{\beta(\as)} f(\as) \,.\nn
\end{align}
The latter are identically zero at lower orders since $g^R(\as)$ and $f(\as)$ start at $\mathcal{O}(\as^4)$ and $\mathcal{O}(\as^3)$ respectively.

The hard function admits a perturbative expansion whose coefficients $H_\kappa^{(n)}$ are defined by
\begin{align}
  \label{eq:hardcoeffs}
H_\kappa(\Phi_1,\mu_H) = \frac{4 \pi \alpha_s (\mu_H)}{4 d_{\kappa_a} d_{\kappa_b}} 
  \sum_{n=0}^\infty \bigg( \frac{\alpha_s(\mu_H)}{4 \pi} \bigg)^n H_\kappa^{(n)}(\Phi_1,\mu_H) \,,
\end{align}
where $d_i$ is the dimension of the colour representation of parton $i$.
Up to N$^3$LL we only need the first two coefficients. 
They can be extracted from the two-loop helicity
amplitudes  calculated in \refcite{Garland:2002ak,Gehrmann:2011ab}, using
the methods described in \refcite{Becher:2013vva}. In addition, we include
the one-loop axial corrections due to the difference between massive top and
massless bottom triangle loops, which were computed in
\refcite{Campbell:2016tcu}. At present, our implementation neglects the
$\ord{\as^3}$ axial contributions to the $q\bar{q}g$ and $qgq$ channels, which
have only been recently calculated in \refcite{Gehrmann:2022vuk}.
Their contributions is expected to be extremely small for the one-jettiness distribution.

We constructed the hard  functions from the known UV- and IR-finite helicity amplitudes for $Z+$jet~\cite{Becher:2013vva,Garland:2002ak,Gehrmann:2011ab}, adding the $Z/\gamma^*$ interference and the decay into massless leptons, producing the final squared matrix elements in an analytical form.
They have been obtained  by
rewriting products of spinor brackets in terms of the kinematic
invariants, writing them in terms of five parity-even invariants and
one parity-odd invariant which is given by the contraction of the Levi-Civita
tensor with four of the external momenta.
Since they are too lengthy to be presented here, we refrain from including
them in the manuscript.
\\~\\

\subsection{$N$-jettiness beam and jet functions}
\label{sec:beam_jet_func}

The beam and jet functions that enter \eq{fact} are the same in the
factorisation formula for every $N$~\cite{Stewart:2010tn}.
The former can be written as convolutions of perturbatively calculable kernels with the standard parton distribution functions (PDFs). 
The beam and jet functions satisfy the RGEs~\cite{Stewart:2010qs, Stewart:2009yx}
\begin{align} \label{eq:beam_rge_mom_space}
\mu\frac{\de}{\de\mu} B_{a}(t, x, \mu)
&= \int \de t^\prime \, \Gamma_B^{a}(t - t^\prime, \mu) B_{a}(t^\prime, x , \mu)
\,, \\ \label{eq:jet_rge_mom_space}
\mu\frac{\de}{\de\mu} J_{c}(s, \mu)
&= \int \de s^\prime \, \Gamma_J^{c}(s - s^\prime, \mu)  J_{c}(s^\prime, \mu)
\,,\end{align}
where $a, c$ can be a quark or a gluon. Formulae for the second beam function are easily obtained by substituting $a\to b$. The anomalous dimensions in \eqs{beam_rge_mom_space}{jet_rge_mom_space} read
\begin{align}
\Gamma_B^{a}(t, \mu)
& = -2 \bigg[ C_{a} \,\Gamma_{\mathrm{cusp}}(\as) \\ & \quad \quad \ \nn + 2 \sum_{R=F,A} D_{aR}\ g^R(\as) \bigg] \cL_0(t, \mu^2)
\\ & \quad \nn    + \gamma^{a}_B(\alpha_s)\, \delta(t)
\,, \\
\Gamma_J^{c}(s, \mu)
&= -2 \bigl[ C_{c} \,\Gamma_{\mathrm{cusp}}(\as) \\ & \quad \quad \ \nn
                                                             + 2 \sum_{R=F,A} D_{cR}\ g^R(\as) \bigr] \cL_0(s, \mu^2)
\\ & \quad \nn    + \gamma^{c}_J(\alpha_s)\, \delta(s)
\,,\end{align}
where we denote the standard plus distributions by~\cite{Ligeti:2008ac}
\begin{align}
\cL_n(x, \mu^m)
= \biggl[ \frac{\theta(x) \ln^n(x/\mu^m)}{x}\biggr]_+
\,,
\end{align}
where $m$ is an integer equal to the mass dimension of $x$.
In order to solve both RGEs we find it convenient to cast \eqs{beam_rge_mom_space}{jet_rge_mom_space} in Laplace space, where momentum convolutions turn into simple products. We denote the Laplace space conjugate functions with a tilde
\begin{align}
	\tilde{B}_{a}(\varsigma_B,x,\mu)  & = \int \de t\,e^{-t/(Q_a e^{\gamma_E} \varsigma_B)} B_{a}(t,x,\mu)\, , \\
	\tilde{J}_{c}(\varsigma_J,\mu)  & =  \int \de s\,e^{-s/(Q_J e^{\gamma_E} \varsigma_J)} J_{c}(s,\mu)\, ,
\end{align}
where the measures $Q_a$ and $Q_J$ are those introduced in the definition of $\Tau_1$ in \eq{tau1def}.
The RGEs for the beam and jet functions can be written as
\begin{widetext}
\begin{align} \label{eq:beam_rge_pos_space}
\mu\frac{\de}{\de\mu} \ln \tilde{B}_{a}(\varsigma_B, x, \mu)
&= -2 \bigl[ C_{a} \,\Gamma_{\mathrm{cusp}}(\as) + 2 \sum_{R=F,A} D_{aR}\ g^R(\as) \bigr] \ln\Bigl(\frac{Q_a \varsigma_B}{\mu^2} \Bigr)
    + \gamma^{a}_B(\alpha_s)
\,, \\ \label{eq:jet_rge_pos_space}
\mu\frac{\de}{\de\mu} \ln \tilde{J}_{c}(\varsigma_J, \mu)
&= -2 \bigl[ C_{c} \,\Gamma_{\mathrm{cusp}}(\alpha_s) + 2 \sum_{R=F,A} D_{cR}\ g^R(\as) \bigr] \ln\Bigl(\frac{Q_J \varsigma_J}{\mu^2} \Bigr)
    + \gamma^{c}_J(\alpha_s)
\,.\end{align}
\end{widetext}
The solutions of \eqs{beam_rge_pos_space}{jet_rge_pos_space} yield the resummed beam and jet functions in Laplace space
\begin{widetext}
\begin{align}
\tilde{B}_{a}(\varsigma_B, x, \mu)
&= \exp\bigl[ 4 C_{a} K_{\Gamma_{\cusp}}(\mu_B,\mu) + 8 \sum_{R=F,A} D_{aR}\  K_{g^R}(\mu_B,\mu) + K_{\gamma^{a}_B}(\mu_B,\mu) \bigr]
\nn \\ & \qquad
\times \tilde{B}(\partial_{\eta_B}, x, \mu_B)\bigg(\frac{Q_a \varsigma_B}{\mu^2_B}\bigg)^{\eta_B}\bigg|_{ \eta_B
   = - 2 [ C_{a} \eta_{\Gamma_{\mathrm{cusp}}}(\mu_B,\mu) + 2 \sum_{R=F,A} D_{aR}\  \, \eta_{g^R}(\mu_B,\mu)] }\,,
\\
\tilde{J}_{c}(\varsigma_J, \mu)
=& \exp\bigl[ 4 C_{c} K_{\Gamma_{\cusp}}(\mu_J,\mu) + 8 \sum_{R=F,A} D_{c R}\  K_{g^R}(\mu_J,\mu) + K_{\gamma^{c}_J}(\mu_J,\mu) \bigr]
\nn \\ & \qquad
\times \tilde{J}(\partial_{\eta_J},\mu_J)\bigg(\frac{Q_J \varsigma_J}{\mu^2_J}\bigg)^{\eta_J}\bigg|_{\eta_J
   = - 2 [ C_{c} \eta_{\Gamma_{\mathrm{cusp}}}(\mu_J,\mu) + 2 \sum_{R=F,A} D_{c R}\  \, \eta_{g^R}(\mu_J,\mu)] }
\,,\end{align}
\end{widetext}
where they are evolved from their canonical scales $\mu_B$ and $ \mu_J$ to an arbitrary scale $\mu$. By performing the inverse Laplace transform, we obtain them in momentum space
\begin{widetext}
\begin{align}
B_{a}(t, x, \mu)
&= \exp\bigl[ 4 C_{a} K_{\Gamma_{\cusp}}(\mu_B,\mu) + 8 \sum_{R=F,A} D_{aR}\  K_{g^R}(\mu_B,\mu) + K_{\gamma^{a}_B}(\mu_B,\mu) \bigr]
\nn \\ & \qquad
\times \tilde{B}(\partial_{\eta_B}, x, \mu_B) \, \frac{e^{-\gamma_E \eta_B}}{\Gamma(\eta_B)} \frac{1}{t} \bigg( \frac{t}{\mu_B^2} \bigg)^{\eta_B}\bigg|_{ \eta_B
   = - 2 [ C_{a} \eta_{\Gamma_{\mathrm{cusp}}}(\mu_B,\mu) + 2 \sum_{R=F,A} D_{aR}\  \, \eta_{g^R}(\mu_B,\mu)] }\,,
\\
J_{c}(s, \mu)
=& \exp\bigl[ 4 C_{c} K_{\Gamma_{\cusp}}(\mu_J,\mu) + 8 \sum_{R=F,A} D_{cR}\  K_{g^R}(\mu_J,\mu) + K_{\gamma^{c}_J}(\mu_J,\mu) \bigr]
\nn \\ & \qquad
\times \tilde{J}(\partial_{\eta_J},\mu_J) \frac{e^{-\gamma_E \eta_J}}{\Gamma(\eta_J)} \frac{1}{s} \bigg( \frac{s}{\mu_J^2} \bigg)^{\eta_J}\bigg|_{\eta_J
   = - 2 [ C_{c} \eta_{\Gamma_{\mathrm{cusp}}}(\mu_J,\mu) + 2 \sum_{R=F,A} D_{c R}\  \, \eta_{g^R}(\mu_J,\mu)] }
\,.\end{align}
\end{widetext}
Similar to the hard functions in \eq{hardcoeffs}, the perturbative components
of the beam and jet functions admit an expansion in terms of powers of the
strong coupling constant and perturbatively calculable coefficients.
For the beam functions these have been recently calculated up to
N$^3$LO~\cite{Stewart:2010qs, Gaunt:2014xga, Gaunt:2014cfa, Ebert:2020unb, Baranowski:2022vcn}
while for the jet functions they have been known for some time~\cite{Bauer:2003pi,Bosch:2004th,Becher:2006qw,Becher:2009th,Becher:2010pd,Bruser:2018rad,Banerjee:2018ozf}.
For our N$^3$LL predictions we only need the beam and jet coefficients up to $\ord{\as^2}$.

\subsection{One-jettiness soft functions}
\label{sec:soft_func}

The soft function for exclusive $N$-jet production was first calculated at NLO in \refcite{Jouttenus:2011wh}.
There, results were presented for the fully differential soft function in $\Tau_N^i$, where $i$ labels
the beam and jet regions, \mbox{$i=a, b, J_1,\ldots,J_N$}. In our case, the NLO soft function appearing in \eq{fact} can be obtained
from these results by specifying $N=1$ and projecting the soft momenta from each region to a single variable,
\begin{align} \label{eq:single_diff_soft_def}
  S^\kappa&(\Tau_1^s, \mu) = \\\nn
  &\int \!\!\de k_a\, \de k_b\, \de k_J \, S_{N=1}^\kappa(\{k_i\},  \mu) \, \delta(\Tau_1^s - k_a - k_b - k_J)
\,,\end{align}
where we have left implicit any angular dependence of the soft functions on
the jet directions, \emph {i.e.} \mbox{$S^\kappa (\Tau_1^s, \mu) \equiv S^\kappa (n_a \cdot
n_J, \Tau_1^s, \mu)$} introduced in \eq{fact} . It satisfies the following RGE
\begin{align} \label{eq:tau1_soft_rge_def}
\mu \frac{\de}{\de\mu} S^{\kappa\,}(\Tau_1^s, \mu)
=\int\!\! \de \ell \, \Gamma_S^\kappa(\Tau_1^s - \ell, \mu) \, S^\kappa(\ell, \mu)
\,,\end{align}
with the anomalous dimensions $\Gamma_S^\kappa(\Tau_1^s, \mu)$ related to those of the fully differential
soft function $S_{N=1}^\kappa(\{k_i\},  \mu)$ by an analogous projection of the soft momentum from each region to a single variable,
\begin{align} \label{eq:single_diff_soft_anomdim}
\Gamma_S^\kappa&(\Tau_1^s, \mu)
= \\\nn& \int \, \de k_a \de k_b \de k_J \, \Gamma^\kappa_{S_{N=1}}(\{k_i\}, \mu) \, \delta(\Tau_1^s -k_a -k_b -k_J)
\,.\end{align}
Explicit expressions for both the fully differential $\Gamma^\kappa_{S_{N=1}}(\{k_i\}, \mu)$ and $S_{N=1}^\kappa(\{k_i\}, \mu)$
at $\ord{\as}$ can be found in \refcite{Jouttenus:2011wh}. Note that in \eqs{single_diff_soft_def}{tau1_soft_rge_def} we have exploited the fact that,
for the present case, the soft function and its anomalous dimensions are trivial matrices in colour space.
The consistency of the factorisation formula, \eq{fact}, implies that the anomalous dimensions of $S^\kappa(\Tau_1^s, \mu)$ can be related to those of the hard, beam, and jet functions by
\begin{align}
\Gamma_S^\kappa(\Tau_1^s, \mu) =& - Q_a \Gamma_B^a(Q_a \Tau_1^s, \mu) -
  Q_b \Gamma_B^b(Q_b \Tau_1^s, \mu) \\& - Q_J \Gamma_J^c(Q_J \Tau_1^s, \mu)
   - 2 {\rm{Re}} \bigl[\Gamma_C^\kappa(\mu) \bigr] \delta(\Tau_1^s)\nn
\,.\end{align}
Using known identities of the plus distributions $\cL_n$~\cite{Ligeti:2008ac}, we find 
\begin{align}
    \label{eq:tau1_soft_anomdim_funcform}
\Gamma_S^\kappa&(\Tau_1^s, \mu)\! =\! 4\! \left[ -\bar{c}^\kappa \Gamma_\cusp(\as)\! +\!\!\!\!\! \sum_{R=F,A} \bar{c}_4^{\kappa,R}
g^R(\as) \right] \!\! \cL_0 (\Tau_1^s, \mu)  \nn \\ \nn &\!\!\!\!\!+ \bigg[                                                \gamma_{S_{N=1}}^\kappa(\as) + 2 \Gamma_\cusp(\as) \bigl(
                           c_s^\kappa L_{ab} + c_t^\kappa L_{ac} + c_u^\kappa
                           L_{bc} \bigr)  \\&\!\!\!\!\!- 2\sum_{R=F, A} g^R(\as) \bigl( c^{\kappa,R}_{4,s} L_{ab} + c^{\kappa,R}_{4,t} L_{bc} + c^{\kappa,R}_{4,u} L_{bc} \bigr) \bigg] \, \delta(\Tau_1^s)
\,,\end{align}
where the noncusp anomalous dimensions of the fully differential soft function~\cite{Jouttenus:2011wh} are given by
\begin{align}
\gamma_{S_{N=1}}^\kappa(\as)
=&-2\sum_{i=a,b,c} \gamma_C^i(\as) - \gamma_B^a(\as) \\ &\nn - \gamma_B^b(\as) - \gamma_J^c(\as) - 2 c_f^\kappa f(\as)
\,,\end{align}
and we use an abbreviated form for the logarithms
\begin{align}\label{eq:kinematic_logs}
L_{ij} \equiv \ln \hat{s}_{ij}
\,,\quad \textrm{with} \quad
\hat{s}_{ij} = \frac{2 q_i \!\cdot\! q_j}{Q_i Q_j}
\,.\end{align}

\Eq{tau1_soft_rge_def} dictates the evolution of the soft function $S^\kappa(\Tau_1^s, \mu)$ from its canonical scale $\mu_S$ to an arbitrary $\mu$.
In addition, it determines its distributional structure in $\Tau_1^s$, up to a boundary term that necessitates explicit computation.
Here, we exploit this in order to solve for the $\ord{\as^2}$ soft function coefficient.
We start by noting that $\Gamma_S^\kappa(\Tau_1^s, \mu)$ in \eq{tau1_soft_anomdim_funcform} has the same distributional form as the zero-jettiness soft function anomalous dimensions~\cite{Stewart:2009yx}.
Thus, we can directly use the known solutions of the zero-jettiness soft function as long as we properly account for the different anomalous dimension coefficients. The logarithmic contributions to the zero-jettiness soft function  were calculated up to N$^3$LO
in \refcite{Billis:2019vxg} and in the following we use the conventions therein. We expand the soft function in momentum space as
\begin{align}
  \label{eq:softcoeffs}
S^\kappa(\Tau_1^s, \mu)
= \sum_{n=0}^\infty \biggl(\frac{\alpha_s(\mu)}{4\pi}\biggr)^n {S}^{\kappa\,(n)}(\Tau_1^s, \mu)
\,,\end{align}
and write the perturbative coefficients in terms of delta functions and plus distributions,
\begin{align}
S^{\kappa\,(m)}(\Tau_1^s, \mu)
= s^{\kappa\,(m)}\, \delta(\Tau_1^s)
+ \sum_{n=0}^{2m-1} S^{\kappa\,(m)}_{n} \, \cL_{n}(\Tau_1^s, \mu)
\,.\end{align}
We find that the $\ord{\as}$ coefficients read
\begin{align}
S^{\kappa\,(1)}_{ 1}
&= -2 (C_{a} + C_{b} + C_{c}) \Gamma_0\nn
\,, \\ 
S^{\kappa\,(1)}_{ 0}\nn
&= - 2 (c_s^\kappa L_{ab} + c_t^\kappa L_{ac} + c_u^\kappa L_{bc}) \Gamma_0
\,,\end{align}
while at $\ord{\as^2}$ they read
\begin{align}
S^{\kappa\,(2)}_{ 3}\nn
&= 2 \Gamma_0^2 (C_{a} + C_{b} + C_{c})^2
  \,,\\ 
S^{\kappa\,(2)}_{ 2}\nn
&= 2 \Gamma_0 (C_{a} + C_{b} + C_{c}) \\ &\qquad \bigl[ \beta_0 + 3\Gamma_0 (c_s^\kappa L_{ab} + c_t^\kappa L_{ac} + c_u^\kappa L_{bc}) \bigr]\nn
                                           \,, \\
  S^{\kappa\,(2)}_{ 1}\nn
&= 4 \Gamma_0^2 \bigl[ (c_s^\kappa L_{ab} + c_t^\kappa L_{ac} + c_u^\kappa L_{bc})^2
    \\ &\qquad \qquad  - \zeta_2 (C_{a} + C_{b} + C_{c})^2\bigr]
\nn \\ & \quad
+ 2 \Gamma_0 \bigl[ 2\beta_0 (c_s^\kappa L_{ab} + c_t^\kappa L_{ac} +
         c_u^\kappa L_{bc}) \nn\\ &\qquad \qquad - (C_{a} + C_{b} + C_{c}) s^{\kappa\,(1)} \bigr]
\nn \\ & \quad\nn
- 2 \Gamma_1 (C_{a} + C_{b} + C_{c})
         \,, \nn
         \\
S^{\kappa\,(2)}_{ 0}
&= 4\Gamma_0^2 (C_{a} + C_{b} + C_{c}) \bigl[ \zeta_3
  (C_{a} + C_{b} + C_{c}) \nn \\ & \qquad \qquad   \nn 
    - \zeta_2 (c_s^\kappa L_{ab} + c_t^\kappa L_{ac} + c_u^\kappa L_{bc})
      \bigr]  \\ \nn & \quad - \gamma^\kappa_{S_{N=1},\,1} - 2\beta_0
                                             s^{\kappa\,(1)}
 \\\nn & \quad - 2 (\Gamma_0 \, s^{\kappa\,(1)} + \Gamma_1) (c_s^\kappa L_{ab} + c_t^\kappa
         L_{ac} + c_u^\kappa L_{bc}) \,.\end{align}
Note that the functions $f(\as)\sim\ord{\as^3}$ and $g^R(\as)\sim\ord{\as^4}$, and therefore they enter in the fixed-order expansion of $S(\Tau_1^s, \mu)$ starting only at N$^3$LL$^\prime$ accuracy.

The boundary terms $s^{\kappa \, (n)}$ are not predicted by the RGE and they
necessitate an explicit computation. At LO they are still trivial,
\begin{align}
s^{\kappa \, (0)} = 1
\,,\end{align}
while at $\ord{\as}$ they have been analytically calculated for arbitrary $N$ and distance measures $Q_i$ in \refcite{Jouttenus:2011wh}.
In the case of one-jettiness they read
\begin{align} \label{eq:nlo_soft_boundary_term}
s^{\kappa\,(1)}
=& 2 c^\kappa_s \bigl[ L_{ab}^2 - \frac{\pi^2}{6} + 2 (I_{ab,c} + I_{ba,c}) \bigr]\nn\\
&+ 2 c^\kappa_t \bigl[ L_{ac}^2 - \frac{\pi^2}{6} + 2 (I_{ac,b} + I_{ca,b}) \bigr]\\
&+ 2 c^\kappa_u \bigl[ L_{bc}^2 - \frac{\pi^2}{6} + 2 (I_{bc,a} + I_{cb,a}) \bigr]\nn
\,,\end{align}
where we use the abbreviation for the finite integrals
\begin{align}
I_{ij,m}
\equiv I_0\Bigl( \frac{\hat{s}_{jm}}{\hat{s}_{ij}}, \frac{\hat{s}_{im}}{\hat{s}_{ij}} \Bigr) \ln\frac{\hat{s}_{jm}}{\hat{s}_{ij}}
+ I_1\Bigl( \frac{\hat{s}_{jm}}{\hat{s}_{ij}}, \frac{\hat{s}_{im}}{\hat{s}_{ij}} \Bigr)
\,,\end{align}
with expressions for $I_{0, 1}(\alpha, \beta)$ given in \refcite{Jouttenus:2011wh}. In our predictions we evaluate \eq{nlo_soft_boundary_term}
for each phase space point on-the-fly in the corresponding reference frame.

The $\mathcal{O}(\alpha_s^2)$ boundary term $s^{\kappa \, (2)}$ was evaluated in \refcite{Boughezal:2015eha,Campbell:2017hsw} in the LAB frame, where the parameters $\rho_i=1$. The result is numeric, and the authors of \refcite{Campbell:2017hsw} provide useful fit functions for the complete NNLO correction for all partonic channels. Nevertheless, in this work, we use a new evaluation of the soft function performed by a subset of the authors of \refcite{BDMR}. This calculation is based on an extension of the \texttt{SoftSERVE} framework \cite{Bell:2018vaa,Bell:2018oqa,Bell:2020yzz} to soft functions with an arbitrary number of light-like Wilson lines. This approach relies on a universal parameterisation of the phase-space integrals, which is used to isolate the singularities of the soft function in Laplace space. The observable-dependent integrations are then performed numerically.

The soft function in the CS frame is then related to that in the LAB frame by a boost along the beam direction. While the invariants $n_i\cdot n_j$ are frame-independent, the soft function implicitly depends on the quantities $\hat{s}_{ij}$ defined in \eq{kinematic_logs}, which are frame-dependent. Specifically, in the LAB and CS frame they are related by
\begin{align}
&\hat{s}^\text{LAB}_{ab}=\hat{s}^\text{CS}_{ab}=1\,,&
&\hat{s}^\text{LAB}_{aJ}=\frac{n_a\cdot n_J}{2}=\rho_a \rho_J \, \hat{s}^\text{CS}_{aJ}\,,
\end{align}
which implies that events with moderately sized $\hat{s}^\text{CS}_{aJ}$ may require us to evaluate the LAB-frame soft function at exceedingly small values of  $\hat{s}^\text{LAB}_{aJ}$, depending on the size of the boost-induced factor $\rho_a \rho_J$. We therefore supplement our numerical calculation with analytic results that can be derived in the asymptotic limit of a jet approaching one of the beam directions, i.e. where $\hat{s}^\text{LAB}_{aJ} \ll 1$ (or $\hat{s}^\text{LAB}_{bJ}\ll1$), to leading power in $\hat{s}^\text{LAB}_{aJ}$ ($\hat{s}^\text{LAB}_{bJ}$) (details will be given in \refcite{BDMR}).

Specifically, we use the symmetry of the soft function under the exchange of the two beam directions to restrict the phase space to configurations with $\hat{s}^\text{LAB}_{aJ} \leq 1/2$. We then divide the phase space into four regions with $\hat{s}^\text{LAB}_{aJ}\leq10^{-12}$, $\hat{s}^\text{LAB}_{aJ}\in [10^{-12},10^{-8}]$, $\hat{s}^\text{LAB}_{aJ}\in [ 10^{-8},10^{-4}]$, and $\hat{s}^\text{LAB}_{aJ}\in [ 10^{-4},1/2]$. In the first region we use the novel analytic leading-power expressions. As power corrections are expected to scale as $\mathcal{O}(\sqrt{\hat{s}^\text{LAB}_{aJ}})$ (modulo logarithms), this means that the accuracy of the leading-power approximation should be at sub-percent level in this region. For the remaining three regions, we construct Chebyshev interpolations of numerical grids, consisting of $4$, $9$ and $43$ sampling points respectively, directly in Laplace space. We construct these interpolations for each interval separately before putting them together.

Following similar considerations as in \sec{beam_jet_func}, we now turn to the resummed soft function in Laplace space which is defined as
\begin{align}
\tilde{S}^\kappa(\varsigma_S,\mu)
= \int \!\!\df \Tau_1^s \, e^{-\Tau_1^s / (e^{\gamma_E} \varsigma_S)} \, S^\kappa(\Tau_1^s,\mu)
\,,\end{align}
and satisfies the multiplicative RGE
\begin{align}
\label{eq:soft_func_rge_laplace_sp}
\mu  \frac{\df}{\df \mu} & \ln  \tilde{S}^\kappa(\varsigma_S,\mu) =  \\
2&\left[ -\bar{c}^\kappa \Gamma_\cusp(\as) +\!\!\!\! \sum_{R=F,A} \bar{c}_4^{\kappa,R}
g^R(\as) \right] \ln \left( \frac{\varsigma_S^2}{\mu^2} \right) \nn \\
+&\,\bigg[\gamma_{S_{N=1}}^\kappa(\as) 
 + 2 \Gamma_\cusp(\as) \bigl(c_s^\kappa L_{ab} + c_t^\kappa L_{ac} + c_u^\kappa L_{bc} \bigr) \, \nn \\
&- 2\sum_{R=F, A} g^R(\as) \bigl( c^{\kappa,R}_{4,s} L_{ab} + c^{\kappa,R}_{4,t} L_{bc} + c^{\kappa,R}_{4,u} L_{bc} \bigr) \bigg] \nn
\,.\end{align}
The solution of \eq{soft_func_rge_laplace_sp} is given by
\begin{widetext}
\begin{align}
\tilde{S}^\kappa(\varsigma_S,\mu)
&= \exp \Bigl\{
2 \bigl(c^\kappa_s L_{ab} +c^\kappa_t L_{ac} + c^\kappa_u L_{bc} \bigr) \eta_{\Gamma_{\mathrm{cusp}}}(\mu_S,\mu)
-2 \sum_{R=F,A}\bigl(c^{\kappa, R}_{4,\, s} L_{ab} + c^{\kappa, R}_{4,t} L_{ac} + c^{\kappa, R}_{4,\, u} L_{bc} \bigr) \eta_{g^R}(\mu_S,\mu)
\nn \\ & \qquad\qquad
+ 4 \, \bar{c}^\kappa K_{\Gamma_{\mathrm{cusp}}}(\mu_S,\mu) - 4 \sum_{R=F,A} \bar{c}^{\kappa, R}_{4} \, K_{g^R}(\mu_S,\mu)+ K_{\gamma^\kappa_S}(\mu_S,\mu)
\Bigr\}
\nn \\ & \qquad\qquad
\times \tilde{S}^\kappa(\partial_{\eta_S},\mu_S) \bigg(\frac{\varsigma_S}{\mu_S}\bigg)^{2 \eta_S}\bigg|_{\eta_S = -2 \, \bar{c}^\kappa \, \eta_{\Gamma_{\mathrm{cusp}}}(\mu_S,\mu)
   + 2 \sum_{R=F,A} \bar{c}^{\kappa, R}_4 \, \eta_{g^R}(\mu_S,\mu)}
\,,\end{align}
and by performing the inverse transform we obtain it in momentum space
\begin{align}
S^\kappa(\Tau_1^s,\mu)
&= \exp \Bigl\{
2 \bigl(c^\kappa_s L_{ab} +c^\kappa_t L_{ac} + c^\kappa_u L_{bc} \bigr) \eta_{\Gamma_{\mathrm{cusp}}}(\mu_S,\mu)
-2 \sum_{R=F,A}\bigl(c^{\kappa, R}_{4,\, s} L_{ab} + c^{\kappa, R}_{4,t} L_{ac} + c^{\kappa, R}_{4,\, u} L_{bc} \bigr) \eta_{g^R}(\mu_S,\mu)
\nn \\ & \qquad\qquad
+ 4 \, \bar{c}^\kappa K_{\Gamma_{\mathrm{cusp}}}(\mu_S,\mu) - 4 \sum_{R=F,A} \bar{c}^{\kappa, R}_{4} \, K_{g^R}(\mu_S,\mu)+ K_{\gamma^\kappa_S}(\mu_S,\mu)
\Bigr\}
\nn \\ & \qquad\qquad
\times \tilde{S}^\kappa(\partial_{\eta_S},\mu_S) \frac{e^{-2\gamma_E \eta_S}}{\Gamma(2\eta_S)} \frac{1}{\Tau_1^s} \biggl( \frac{\Tau_1^s}{\mu_S} \biggr)^{2\eta_S}
\bigg|_{\eta_S = -2 \, \bar{c}^\kappa \, \eta_{\Gamma_{\mathrm{cusp}}}(\mu_S,\mu)
   + 2 \sum_{R=F,A} \bar{c}^{\kappa, R}_4 \, \eta_{g^R}(\mu_S,\mu)}
\,.\end{align}

\subsection{Final resummed and matched formulae}
\label{sec:resummed_formula}


Combining all the previous ingredients together and using the following
definitions
\begin{align}
K_{\gamma_{\rm tot}} &=   -2 n_g  K_{\gamma^g_C}(\mu_S, \mu_H) + 2 (n_g-3) K_{\gamma^q_C}(\mu_S, \mu_H)   - (n_g-n^{\kappa_J}_g) K_{\gamma^g_J}(\mu_J, \mu_B) - n_g K_{\gamma^g_J}(\mu_S, \mu_J) \nn \\
& \quad  + (n_g-2-n^{\kappa_J}_g) K_{\gamma^q_J}(\mu_J, \mu_B)  + (n_g-3) K_{\gamma^q_J}(\mu_S, \mu_J)  +  2 c^\kappa_f K_f(\mu_H,\mu_S)\,, 
\end{align}
where $n_g$ is the total number of gluons and $n^{\kappa_J}_g$ the number of
gluons in the final state, we arrive at the 
resummation formula which, when evaluated at N$^3$LL accuracy, reads
\begin{align}
\label{eq:n3llformula}
\frac{\de\sigma^{\textsc{n}^3\textsc{ll}}}{\de\Phi_1\de\Tau_1}  = & \sum_{\kappa} 
\exp\bigg\{
4(C_{a}+C_{b})
  K_{\Gamma_{\cusp}}(\mu_B, \mu_H) + 4 C_{c}
  K_{\Gamma_{\cusp}}(\mu_J, \mu_H) 
  - 2 (C_{a}+C_{b}
  + C_{c})
  K_{\Gamma_{\cusp}}(\mu_S, \mu_H) \nn \\&\quad
  \qquad \quad - 2 C_{c} L_J\ \eta_{\Gamma_{\cusp}}(\mu_J, \mu_H)
  -2 ( C_{a} L_B 
  + C_{b} L'_B ) \eta_{\Gamma_{\cusp}}(\mu_B, \mu_H)
   + K_{\gamma_{\rm tot}} \nn \\&\quad
  + \bigg[ 
  C_{a} \ln\left(\frac{Q^2_a u }{s t} \right)
  + C_{b} \ln\left( \frac{Q^2_b t }{s u }\right) 
  + C_{\kappa_j} \ln\left( \frac{Q^2_J s }{t u }\right)  
  +  (C_{a}+C_{b}+C_{c}) L_S \bigg] \eta_{\Gamma_{\cusp}}(\mu_S, \mu_H) 
  \nn \\
& +  \sum_{R=F,A}\bigg[ 8\, \big(D_{aR}+D_{bR}\big) K_{g^R}(\mu_B,\mu_H) + 8\,  D_{cR} K_{g^R}(\mu_J,\mu_H) \, \nn \\
&\qquad - 4\,  \big(D_{aR}+D_{bR}+D_{cR}\big) K_{g^R}(\mu_S,\mu_H) \,  - 4 D_{cR} L_J  \eta_{g^R}(\mu_J,\mu_H)\,  - 4\, \big(D_{aR} L_B + D_{bR} L^\prime_B\big) \eta_{g^R}(\mu_B,\mu_H) \nn \\
& \qquad + 2\, \bigg[D_{aR} \ln \bigg(\frac{Q^2_a u}{s t}\bigg)+D_{bR} \ln
  \bigg(\frac{Q^2_b t}{s u}\bigg) + D_{cR} \ln \bigg(\frac{Q^2_J s}{t
  u}\bigg)\, + \big(D_{aR}+D_{bR}+D_{cR}\big)
  L_S\bigg]\eta_{g^R}(\mu_S,\mu_H)\bigg] \bigg\}
                                                                    \nn\\
&
  \times  H_{\kappa}(\Phi_1, \mu_H) \tilde{S}^\kappa \big(\partial_{\eta_S} + L_S,
  \mu_S\big)
\tilde{B}_{\kappa_a}(\partial_{\eta_B} + L_B,x_a,\mu_B) \tilde{B}_{\kappa_b}(\partial_{\eta_B'}+L_B',x_b,\mu_B)\,\tilde{J}_{\kappa_J}(\partial_{\eta_J}+L_J,\mu_J)
  \nn\\& \quad \times
\frac{Q^{-\eta_{\rm tot}}}{{\Tau_1}^{1-\eta_{\rm tot}}}\, \frac{\eta_{\rm tot}\ e^{-\gamma_E \eta_{\rm tot}}}{\Gamma(1+\eta_{\rm tot})} \, ,
\end{align}
\end{widetext}
where the terms
\begin{align}
\eta_S & = -2 \, \bar{c}^\kappa \, \eta_{\Gamma_{\mathrm{cusp}}}(\mu_S,\mu) + 2 \sum_{R=F,A} \bar{c}^{\kappa, \, R}_4 \, \eta_{g^R}(\mu_S,\mu)\, ,\nn\\
\eta_B &= - 2 \big[ C_{a} \eta_{\Gamma_{\mathrm{cusp}}}(\mu_B,\mu) + 2
         \sum_{R=F,A} D_{aR} \, \eta_{g^R}(\mu_B,\mu) \big]\,, \nn
  \end{align}
\begin{align}
\eta^\prime_B & = - 2 \big[ C_{b} \eta_{\Gamma_{\mathrm{cusp}}}(\mu_B,\mu) + 2 \sum_{R=F,A}
                D_{bR} \, \eta_{g^R}(\mu_B,\mu) \big]  \,, \nn \\
\eta_J &= - 2 \big[ C_{\kappa_c} \eta_{\Gamma_{\mathrm{cusp}}}(\mu_J,\mu) + 2 \sum_{R=F,A} D_{cR} \, \eta_{g^R}(\mu_J,\mu) \big] \, , \nn
\end{align}
are combined as  
\begin{align}
\eta_{\mathrm{tot}} &= \eta_B+\eta_{B}^\prime+\eta_J+ 2\eta_S\, ,\nn
\end{align}
and we have also introducted the  definitions 
\begin{align}
L_H & = \ln \left(\frac{Q^2}{\mu_H^2} \right)\,,\ \nn
L_B = \ln \left(\frac{Q_a Q}{\mu_B^2} \right)\,,\ 
      L'_B = \ln \left(\frac{Q_b Q}{\mu_B^2} \right)\,,\\
 L_J &= \ln \left(\frac{Q_J Q}{\mu_J^2} \right)\,,\
L_S = \ln \left(\frac{Q^2}{\mu_S^2} \right)\,.\nn
\end{align}
In the previous equation all the $K_X$ and $\eta_X$ evolution functions are
evaluated at N$^3$LL accuracy and the  boundary terms of the hard, soft, beam and jet
functions in the second to last line are
implicitly expanded up to relative $\ord{\as^2}$.
The complete formula with the boundary terms expanded out is presented in
\app{resn3ll}. 

While Sudakov logarithms at small $\Tau_1$ invalidate the perturbative convergence
and call for their resummation at all orders, as $\Tau_1$ approaches the hard scale
they are no longer considered large. In this regime, the spectrum is correctly described
by fixed-order predictions. In addition, $\Tau_1$ is subject to
the constraint ${\Tau_1/\Tau_0 \leq 1-1/N}$, with $N\!=\!2 \, (N\!=\!3)$ at NLO (NNLO).
Therefore, in order to achieve a proper description throughout
the $\Tau_1$ spectrum while satisfying the $\Tau_1/\Tau_0$ constraint,
we construct two-dimensional (2D) profile scales that modulate the transition to the FO region
as a function of both $\Tau_1/\mu_\FO$ and $\Tau_1/\Tau_0$, with $\mu_\FO$
the fixed-order scale.
These profile scales correctly implement the phase space constraint in $\Tau_1/\Tau_0$,
reducing to $\Tau_1$-dependent profile scales when it is satisfied and asymptoting to
$\mu_\FO$ when it is violated. A detailed discussion of our 2D profile scale
construction is given in \sec{2d_profiles}.

A reliable theoretical prediction must include a thorough uncertainty estimate by
exploring the entire space of possible scale variations. In our analysis, we achieve this
by means of $\Tau_1$ profile scale variations, see
e.g. \refcite{Alioli:2019qzz}. Specifically, our final uncertainty
is obtained by separately estimating the uncertainties related to resummation
and the FO perturbative expansion.
Since these are considered to be uncorrelated, we sum them in quadrature.

In order to achieve a valid description also in the tail region of the
one-jettiness distribution, this resummed result is matched  to the NLO predictions for $\gamma^*/Z+2$~jets
production (NLO$_2$), using a standard additive matching prescription
\begin{align}
  \label{eq:matched}
  \frac{\de\sigma^{\mathrm{N}^3\mathrm{LL+NLO}_2}}{\de\Phi_1\de\Tau_1}  = &
  \frac{\de\sigma^{\mathrm{N}^3\mathrm{LL}}}{\de\Phi_1\de\Tau_1} +
                                                                               \frac{\de\sigma^{\mathrm{Nons.}}}{\de\Phi_1\de\Tau_1} \,,
  \\
  \frac{\de\sigma^{\mathrm{Nons.}}}{\de\Phi_1\de\Tau_1}   =  &
                                                              \left( \frac{\de\sigma^{\mathrm{NLO}_2}}{\de\Phi_1\de\Tau_1}                                                                 - \left.  \frac{\de\sigma^{\mathrm{N}^3\mathrm{LL}}}{\de\Phi_1\de\Tau_1} \right|_{\ord{\as^2}}\right) \theta (\Tau_1) \,,\nn
\end{align}
where the last term of the second equation above is the NNLO singular contribution. Similar formulae readily apply at lower orders.
The NLO predictions  for $Z/\gamma^*$+2 jets are obtained
from \geneva, which implements a local FKS
subtraction~\cite{Frixione:1995ms}, using tree-level and one-loop amplitudes from
\openloopsTwo~\cite{Buccioni:2019sur}.

We note that in \eq{matched} we have written the highest accuracy as $\mathrm{N}^3\mathrm{LL+NLO}_2$ because
for the plots presented in this paper we are focusing on the $\Tau_1$ spectrum above a finite value $\Tau_1 > 0$.
Removing the $\as^3\, \delta(\Tau_1)$ contribution which is present in the
singular term but is missing in the $\mathrm{NLO}_2$ differential
cross section, the formula in  \eq{matched}  can be extended to achieve
$\mathrm{N}^3\mathrm{LL+NNLO}_1$ accuracy for quantities integrated over
$\Tau_1$. 

We also note that there is some freedom when evaluating  $\Tau_1$ on events with two or three partons. In this work, we
use $N$-jettiness as a jet algorithm \cite{Stewart:2015waa} and minimise over all possible jet directions $n_J$ obtained by an exclusive clustering procedure $\widetilde{\Tau_1} = \min_{n_J} \Tau_1$.
This means that we recursively cluster together emissions in the
\mbox{$E$-scheme} using the $\Tau_1$ metric in \eq{tau1def} until we are left
with exactly one jet. The resulting jet is then made massless by rescaling its
energy to match the modulus of its three-momentum; the jet direction is then taken to be $\vec n_J$.
We stress that this choice is intrinsically different from  determining the jet axis a priori by employing an inclusive jet clustering,  as done for example in refs.~\cite{Boughezal:2015dva,Boughezal:2015ded,Boughezal:2015aha,Campbell:2019gmd}.

This difference has also the interesting consequence that one has to be careful when defining $\widetilde{\Tau_1}$ via the exclusive jet clustering procedure in a frame which depends on the jet momentum. There are indeed choices of the clustering metric that render the $\widetilde{\Tau_1}$ variable  so defined infrared (IR) unsafe. A particular example is given by  the  frame where the system of the colour-singlet and the jet has zero rapidity $Y_{LJ}=0$ (underlying-Born frame) which was instead previously studied for the inclusive jet definition~\cite{Campbell:2019gmd}. A detailed discussion of these features and a comparison of the  size of nonsingular power corrections for these alternative $\Tau_1$ definitions is beyond the scope of this work and will be presented elsewhere.

\section{Numerical Implementation and Results}
\label{sec:implementation}
\begin{figure*}[ht!]
  \begin{subfigure}[b]{\rescaletwoplots}
    \includegraphics[width=\textwidth]{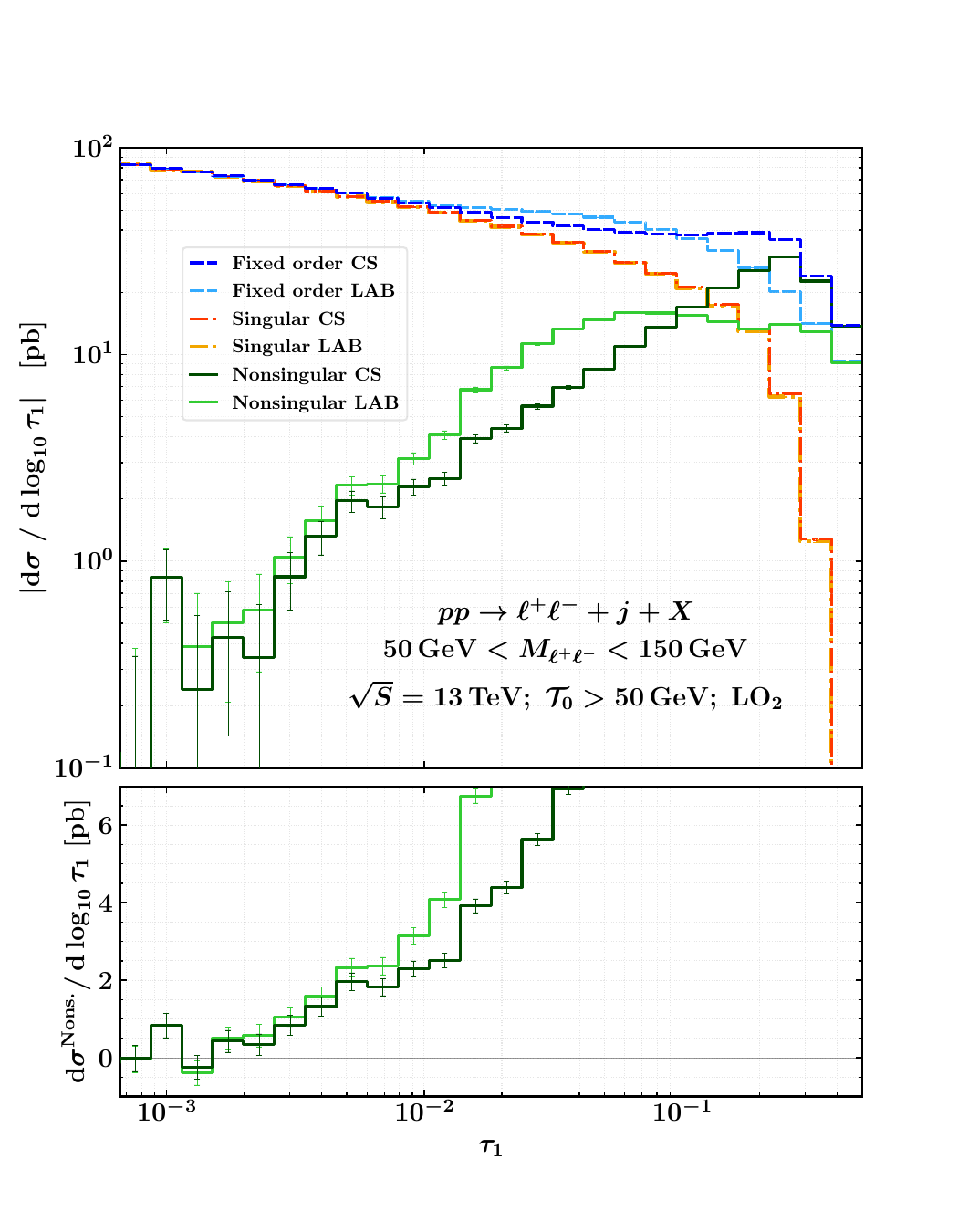}%
  \end{subfigure}
  \hspace*{\hspacebetweentwoplots}
  \begin{subfigure}[b]{\rescaletwoplots}
    \includegraphics[width=\textwidth]{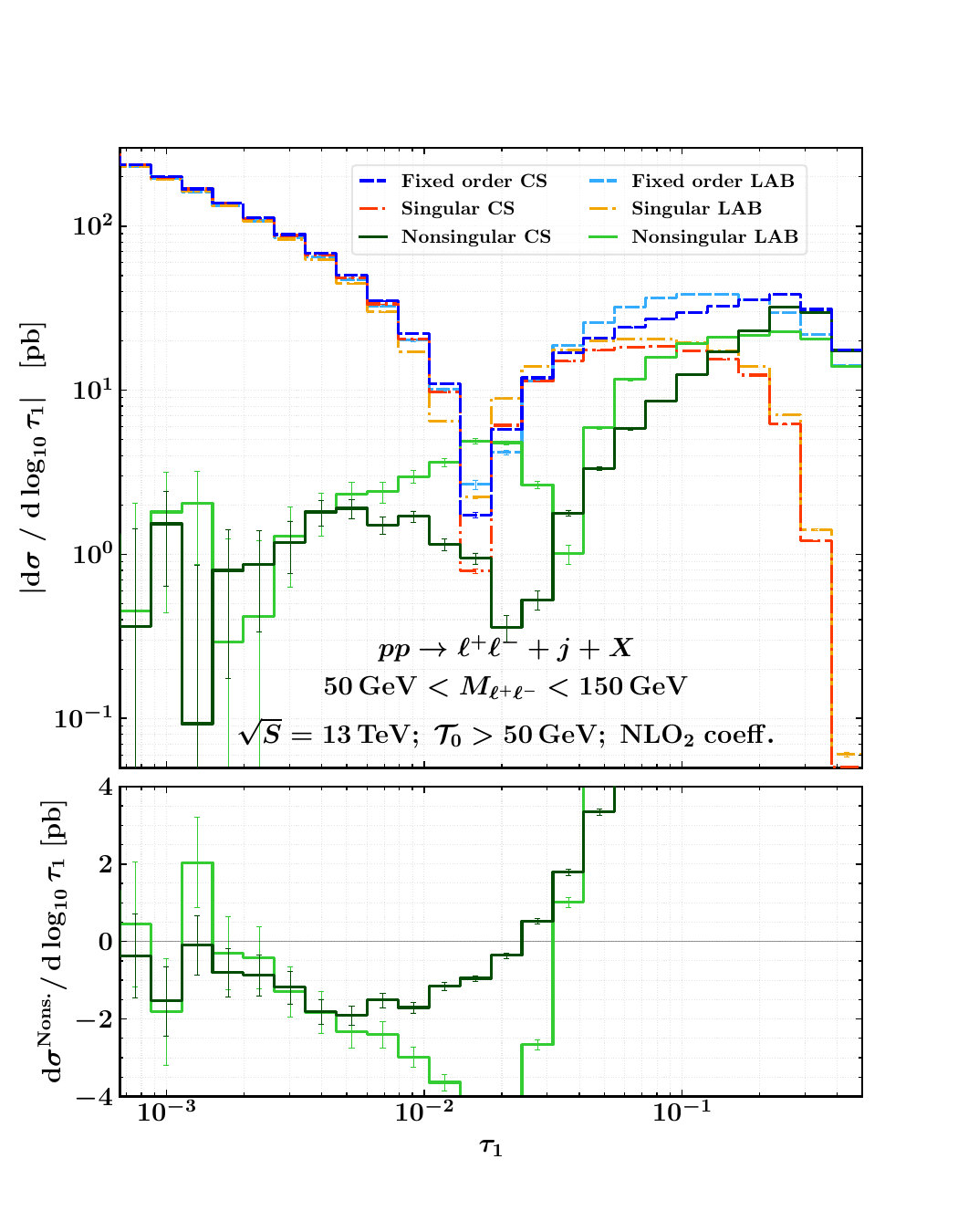}%
  \end{subfigure}
  \vspace{\spaceabovefigurecaption}
  \caption{Absolute values of the $\tau_1 =  \Tau_1 / m_{T}$ spectra with
    $\Tau_0 > 50$~GeV for fixed-order, singular and nonsingular contributions
    at $\ord{\as^2}$ (left) and at pure $\ord{\as^3}$ (right) on a logarithmic
    scale (upper frames) and signed values for the nonsingular on a linear scale (lower frames).
  Results for both the laboratory frame (LAB) and the frame where the
  colour-singlet system has zero rapidity (CS) are shown. Statistical errors from Monte Carlo integration, shown as thin vertical error bars, become sizeable at extremely low $\tau_1$ values.
    \label{fig:nonsing_Tau0_50}
  }
\end{figure*}
\begin{figure*}[ht!]
  \begin{subfigure}[b]{\rescaletwoplots}
    \includegraphics[width=\textwidth]{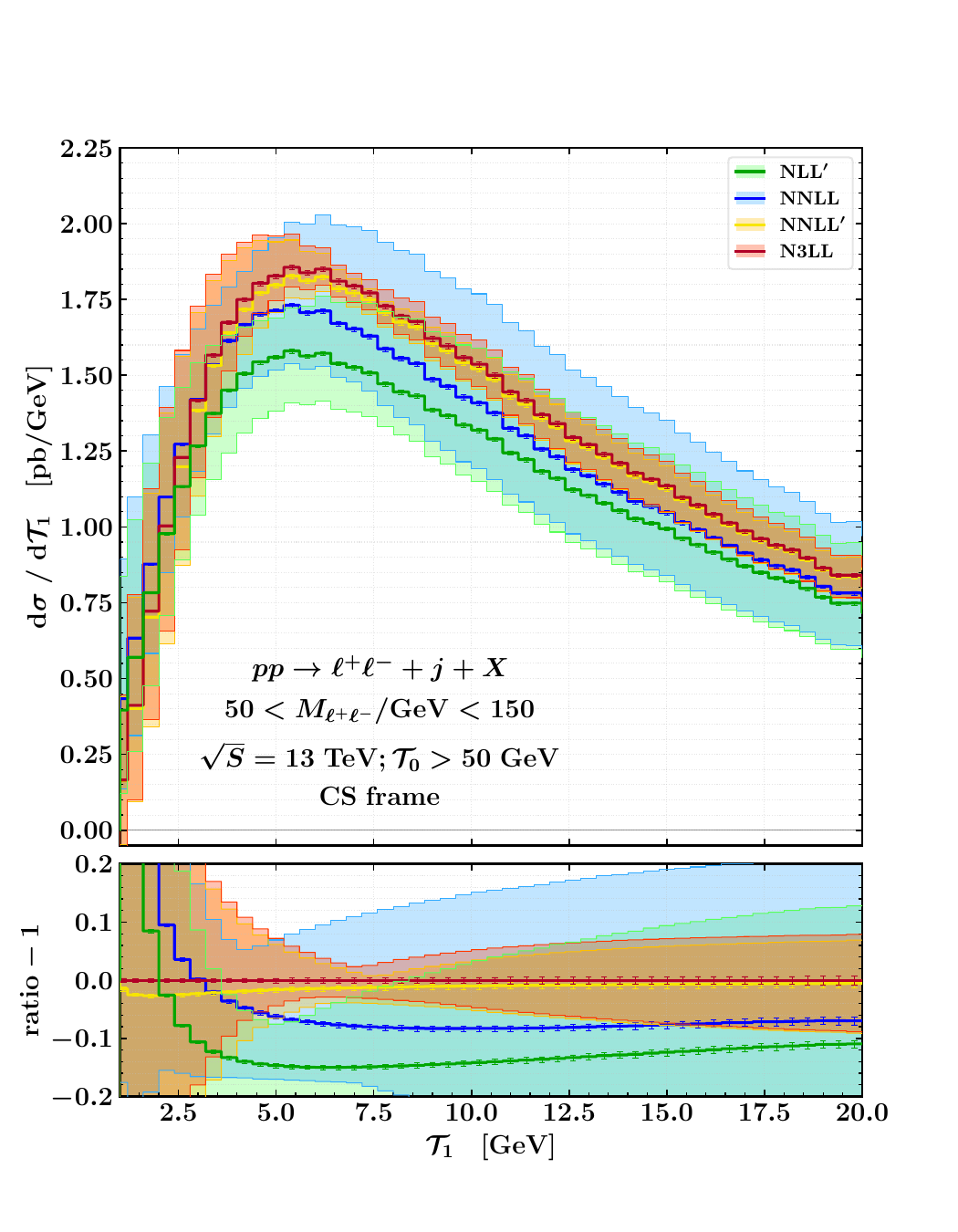}%
  \end{subfigure}
  \hspace*{\hspacebetweentwoplots}
  \begin{subfigure}[b]{\rescaletwoplots}
    \includegraphics[width=\textwidth]{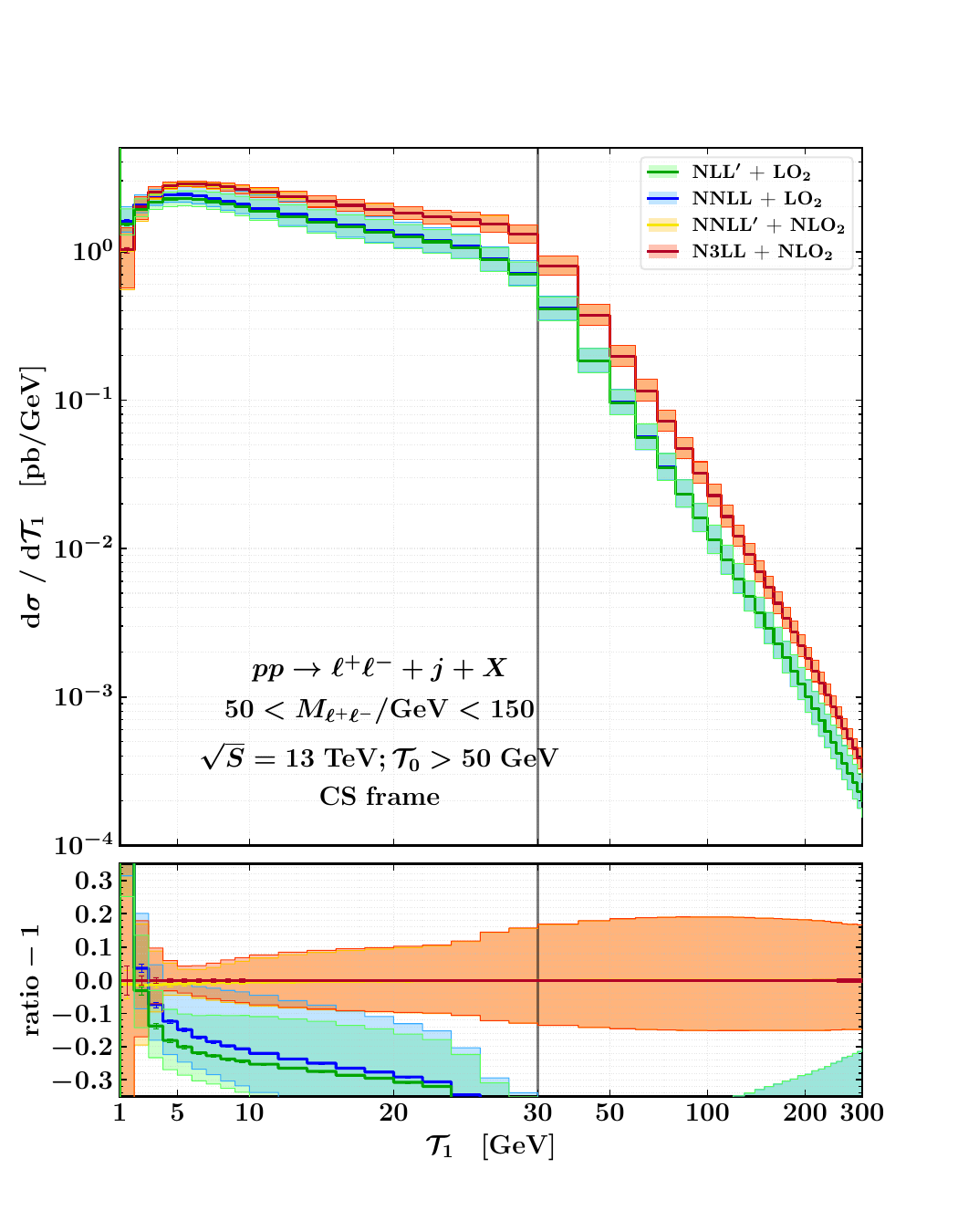}%
  \end{subfigure}
  \vspace{\spaceabovefigurecaption}
  \caption{Resummed (left) and matched (right) results for one-jettiness distribution with $\Tau_0 > 50$~GeV.}
    \label{fig:res_match_Tau0_50}
\end{figure*}
\begin{figure*}[ht!]
  \begin{subfigure}[b]{\rescaletwoplots}
    \includegraphics[width=\textwidth]{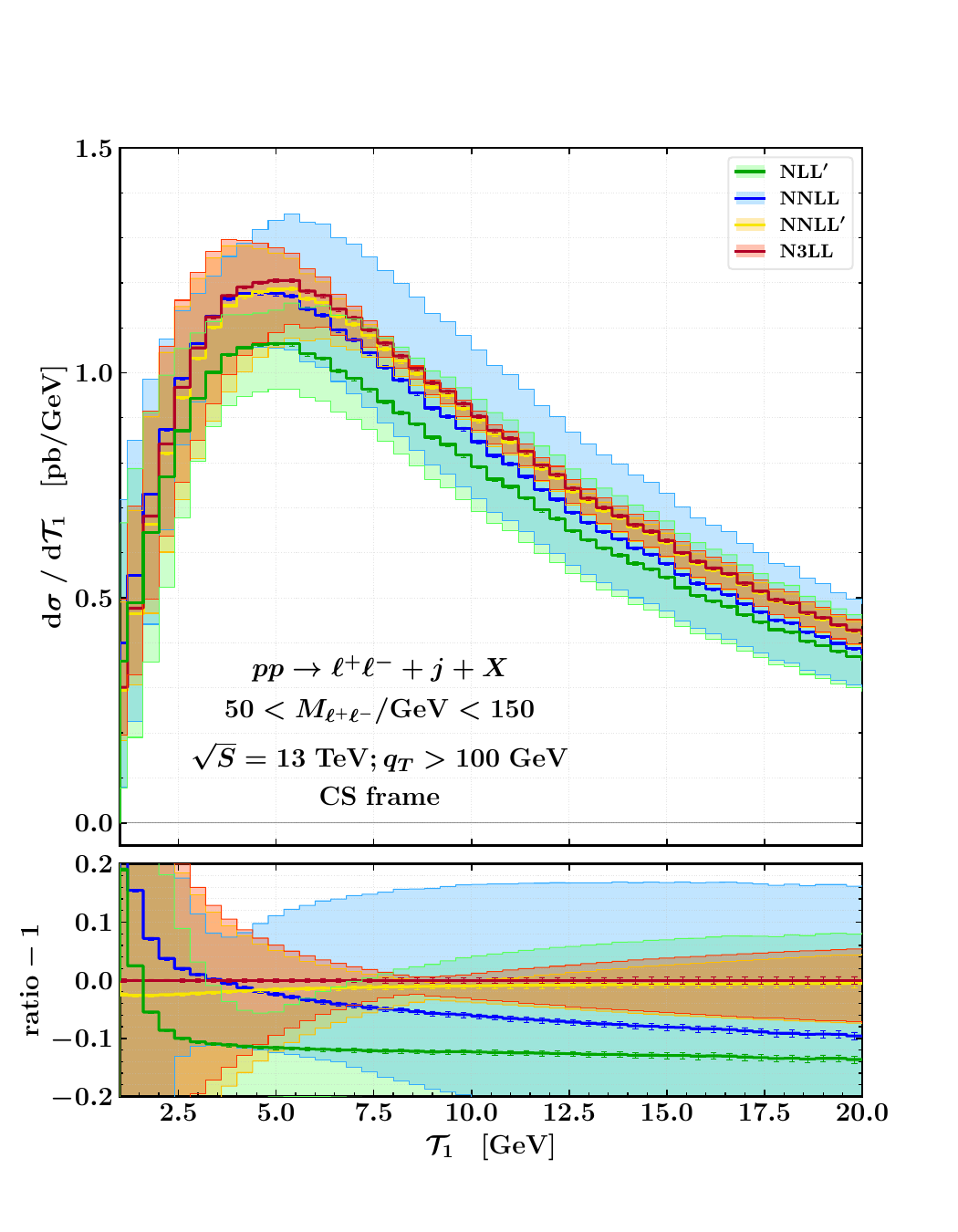}%
  \end{subfigure}
  \hspace*{\hspacebetweentwoplots}
  \begin{subfigure}[b]{\rescaletwoplots}
    \includegraphics[width=\textwidth]{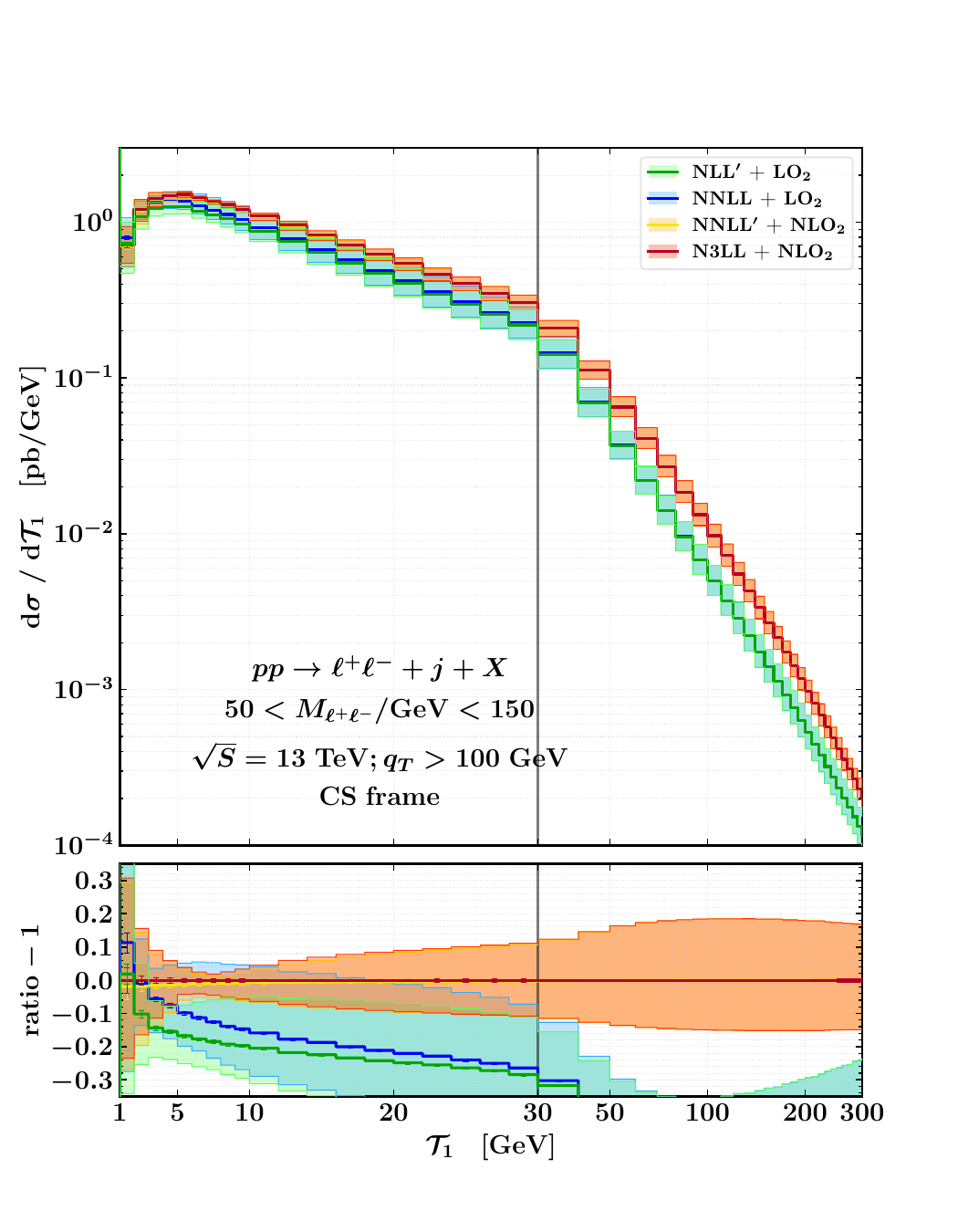}%
  \end{subfigure}
  \vspace{\spaceabovefigurecaption}
  \caption{Resummed (left) and matched (right) results for one-jettiness distribution with $q_T > 100$~GeV.}
    \label{fig:res_match_qT_100}
\end{figure*}
\begin{figure*}[ht!]
	\begin{subfigure}[b]{\rescaletwoplots}
		\includegraphics[width=\textwidth]{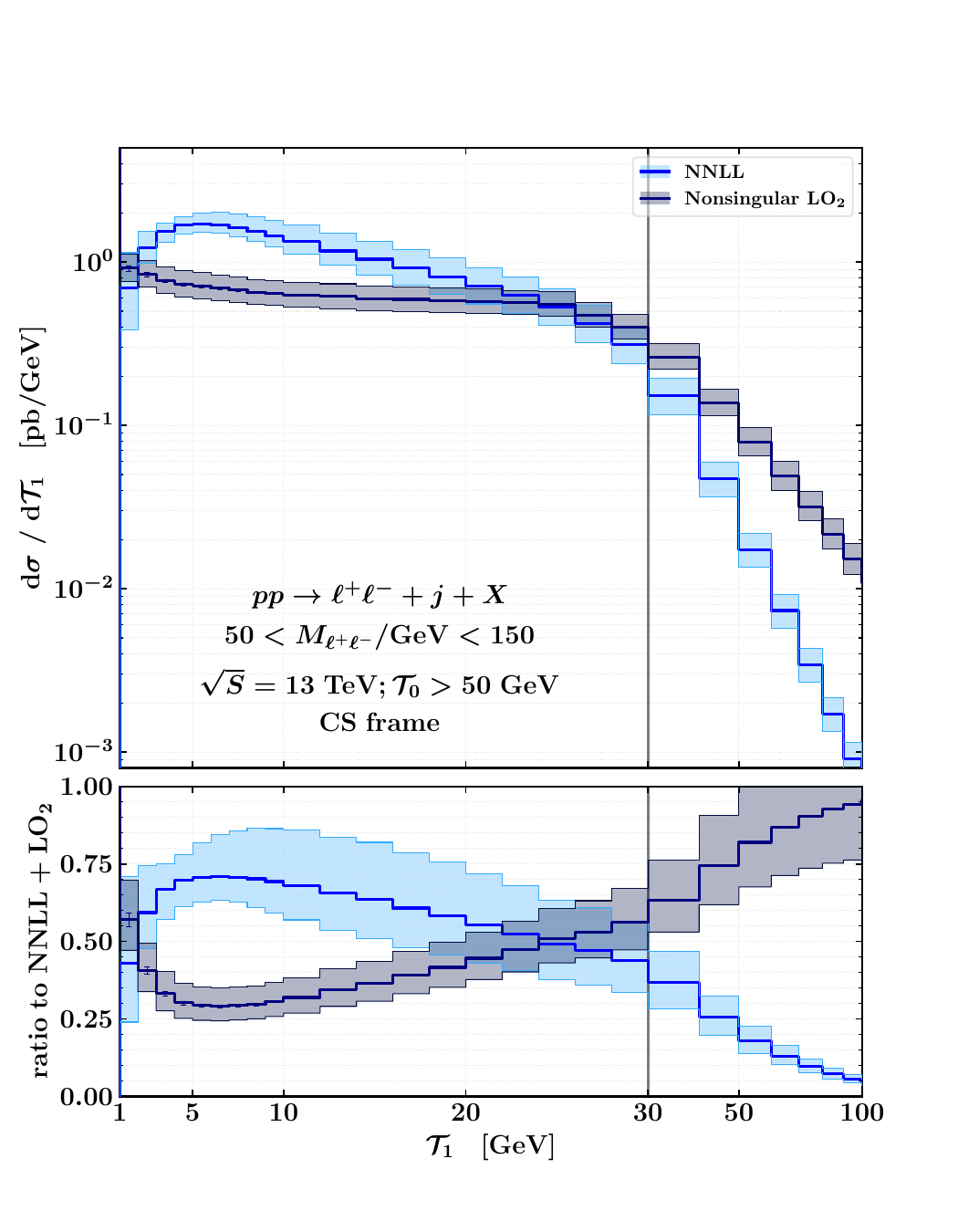}%
	\end{subfigure}
	\hspace*{\hspacebetweentwoplots}
	\begin{subfigure}[b]{\rescaletwoplots}
		\includegraphics[width=\textwidth]{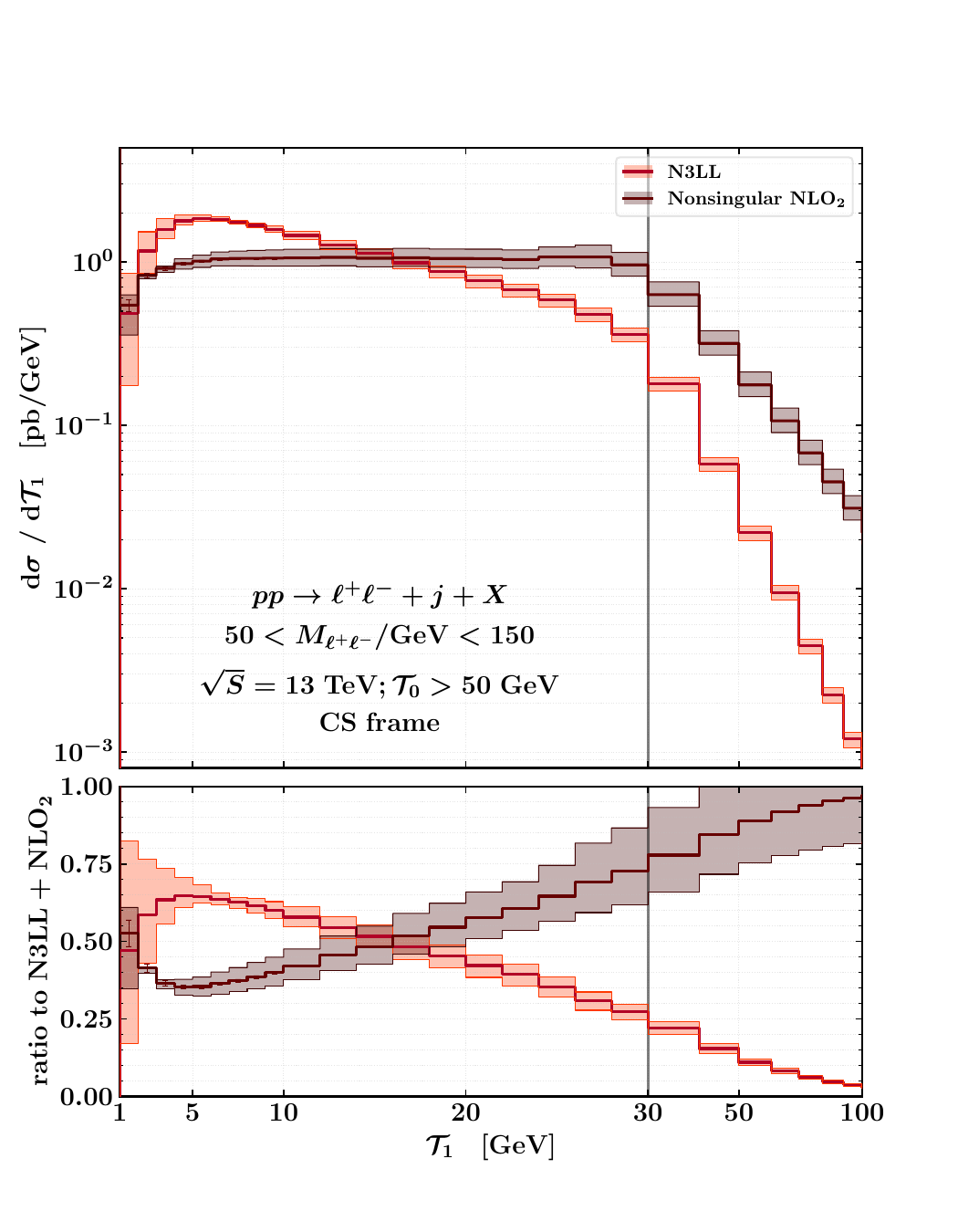}%
	\end{subfigure}
	\vspace{\spaceabovefigurecaption}
	\caption{Comparison between resummed and nonsingular contributions at
		NNLL+LO$_2$ (left) and N$^3$LL+NLO$_2$ (right) for one-jettiness
		distribution with $\Tau_0 > 50$~GeV. The lower inset shows the ratio to
		the corresponding matched prediction.}
	\label{fig:res_nons_Tau0_50}
\end{figure*}
\begin{figure*}[ht!]
\begin{minipage}[c]{0.49\textwidth}
    \includegraphics[width=\textwidth]{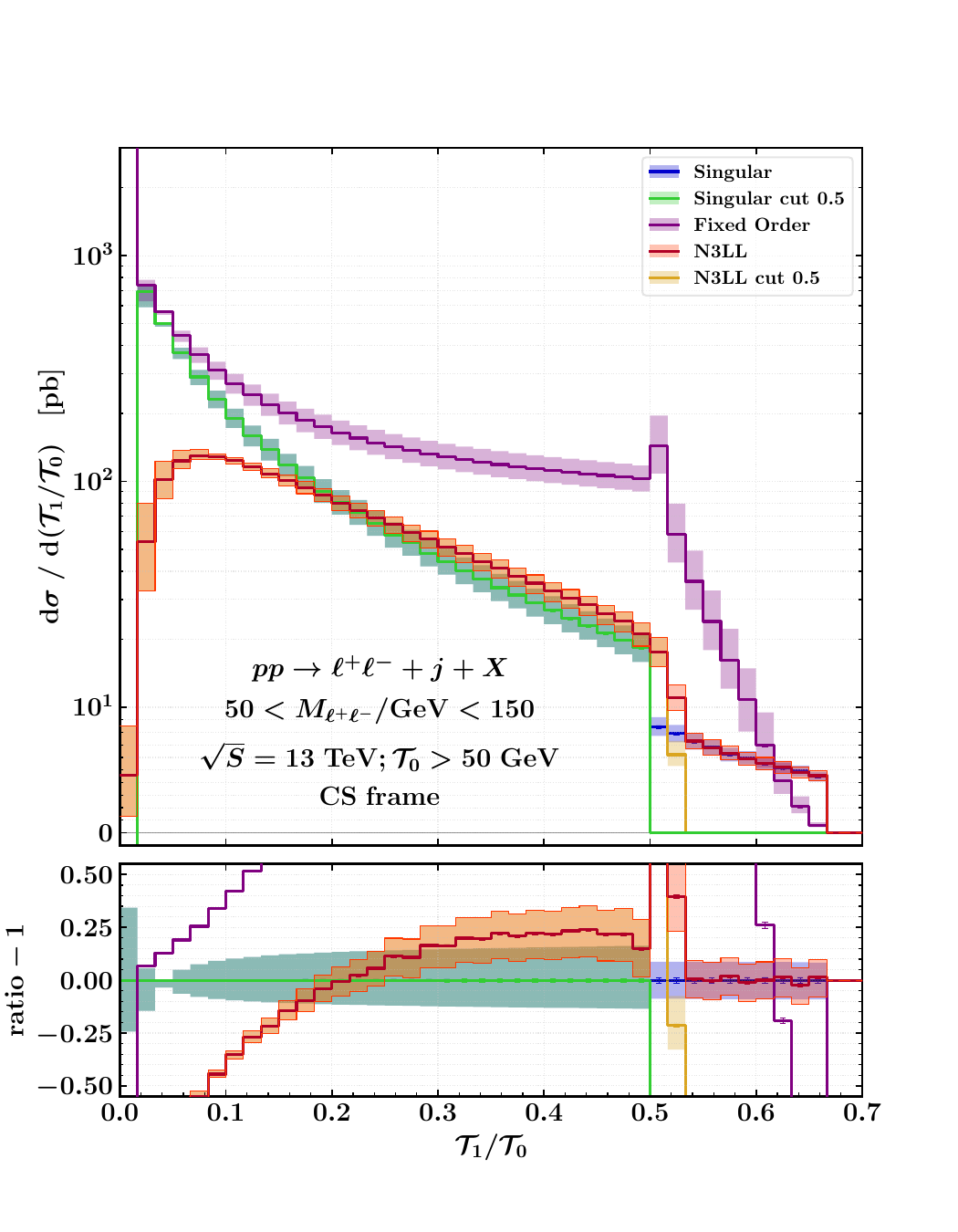}%
  \vspace{\spaceabovefigurecaption}
  \caption{Comparisons between resummed N$^3$LL results, fixed-order NLO$_2$ and singular ones for  $\Tau_0 > 50$~GeV with or without a hard cut at $\Tau_1 / \Tau_0 < 0.5$ on the $\ord{\as^3}$ singular contribution.}
  \label{fig:tau1t_cuts_Tau0_50}
\end{minipage}
  \begin{minipage}[c]{0.49\textwidth}
  \includegraphics[width=\textwidth]{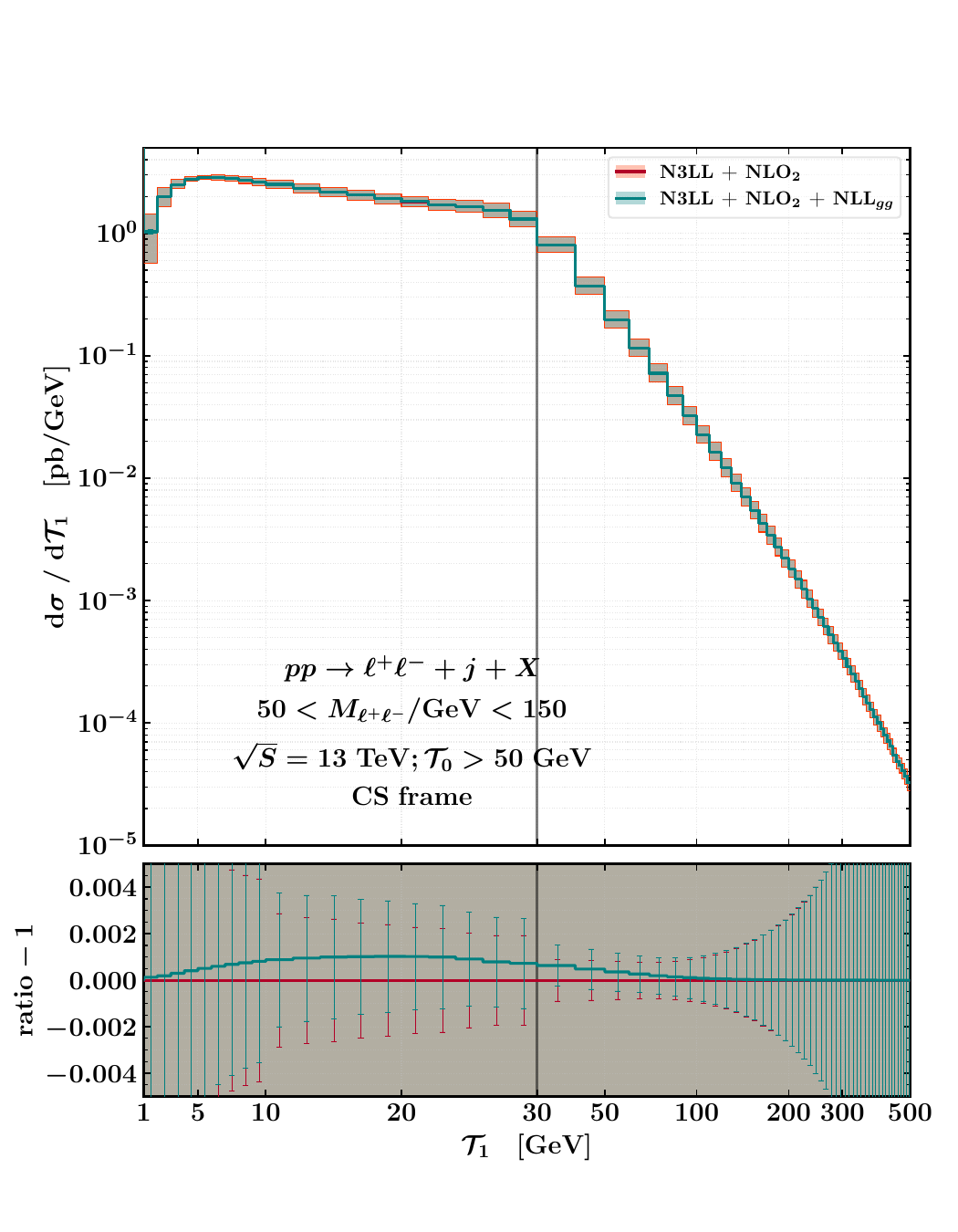}%
  \vspace{\spaceabovefigurecaption}
  \caption{Effects of the inclusion of the NLL resummation of the $gg$ loop-induced channel on top of the N$^3$LL+NLO$_2$ matched predictions.
   \phantom{-------------------------------------------------------------------}
    \label{fig:matched_gg_Tau0_50}
  }
  \end{minipage}
\end{figure*}
\begin{figure*}[ht!]
	\begin{subfigure}[b]{\rescaletwoplots}
		\includegraphics[width=\textwidth]{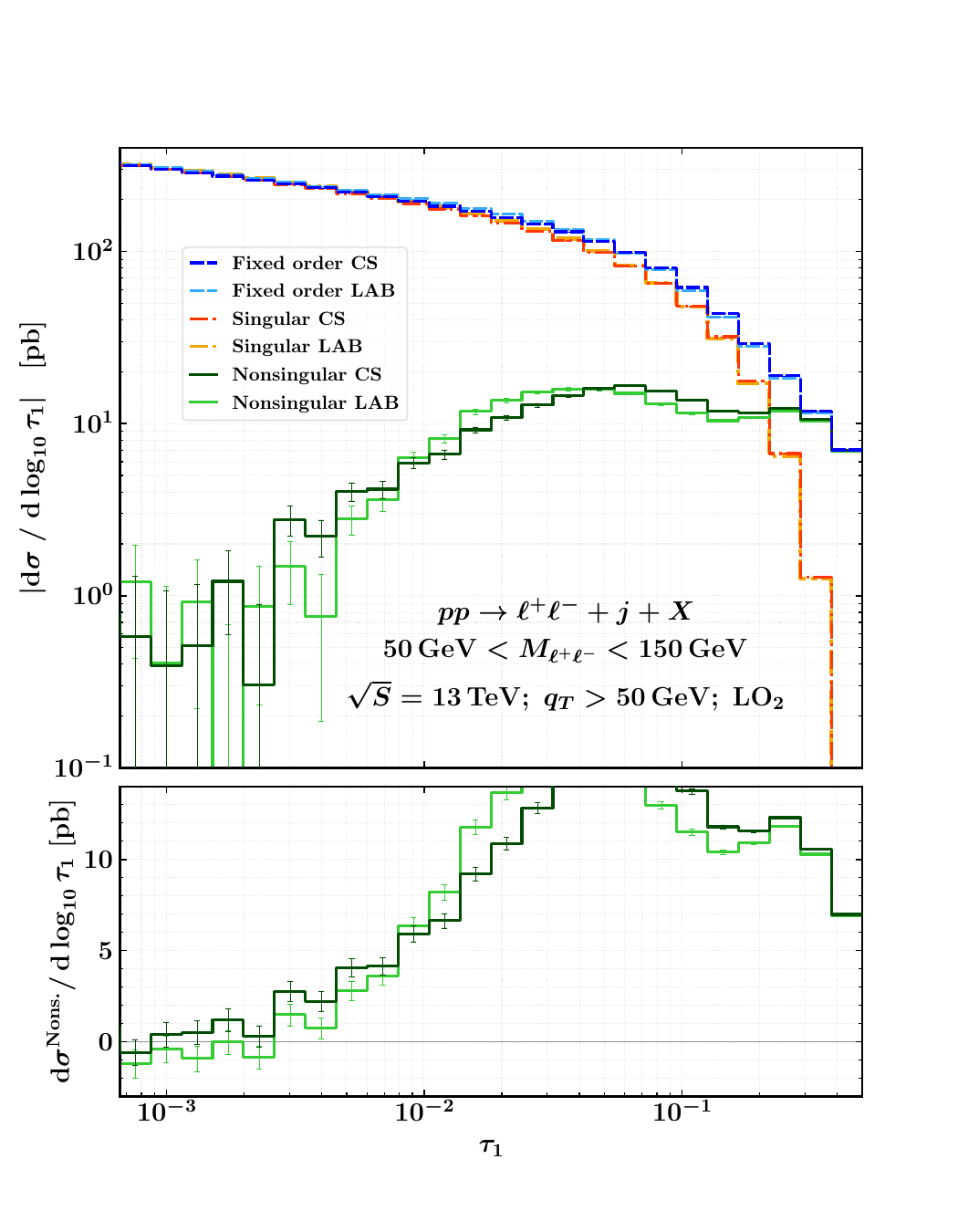}%
	\end{subfigure}
	\hspace*{\hspacebetweentwoplots}
	\begin{subfigure}[b]{\rescaletwoplots}
		\includegraphics[width=\textwidth]{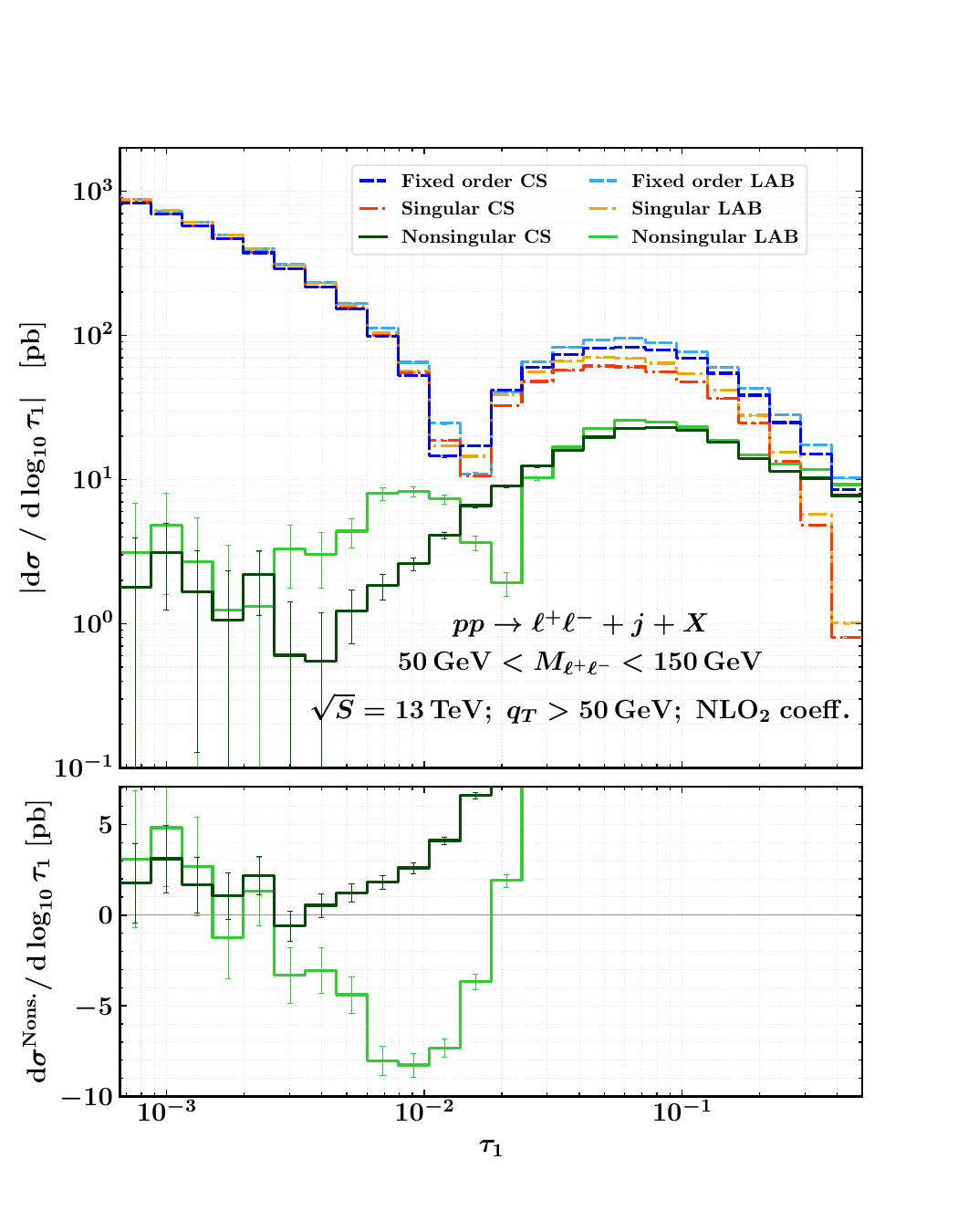}%
	\end{subfigure}
	\vspace{\spaceabovefigurecaption}
	\caption{Absolute values of the $\tau_1 =  \Tau_1 / m_{T}$ spectra with
    $q_T > 50$~GeV for fixed-order, singular and nonsingular contributions
    at $\ord{\as^2}$ (left) and at pure $\ord{\as^3}$ (right) on a logarithmic
    scale (upper frames) and signed values for the nonsingular on a linear scale (lower frames).
  Results for both the laboratory frame (LAB) and the frame where the
  colour-singlet system has zero rapidity (CS) are shown. Statistical errors from Monte Carlo integration, shown as thin vertical error bars, become sizeable at extremely low $\tau_1$ values.
		\label{fig:nonsing_qT_50}
	}
\end{figure*}
\begin{figure*}[ht!]
	\begin{subfigure}[b]{\rescaletwoplots}
		\includegraphics[width=\textwidth]{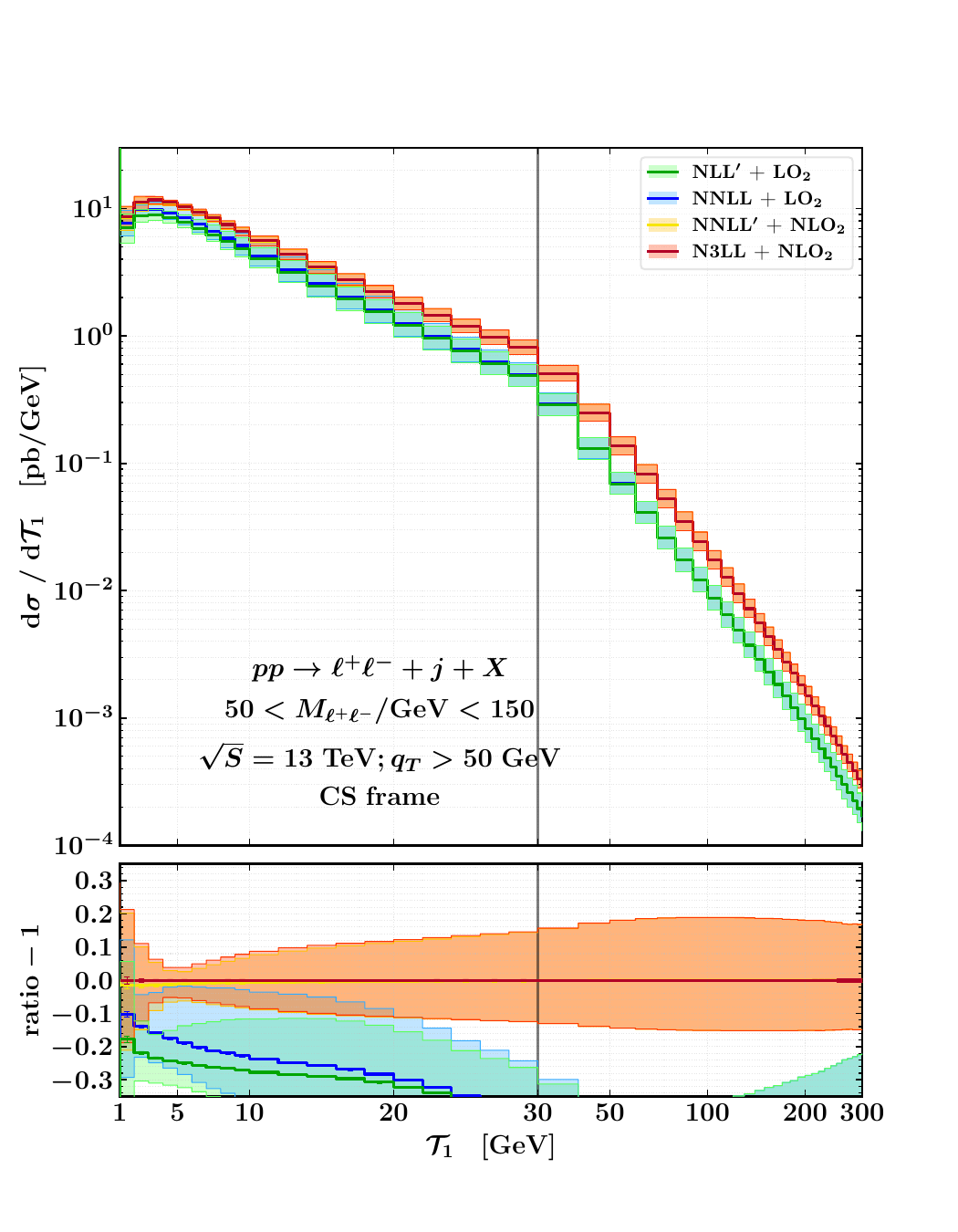}%
	\end{subfigure}
	\hspace*{\hspacebetweentwoplots}
	\begin{subfigure}[b]{\rescaletwoplots}
		\includegraphics[width=\textwidth]{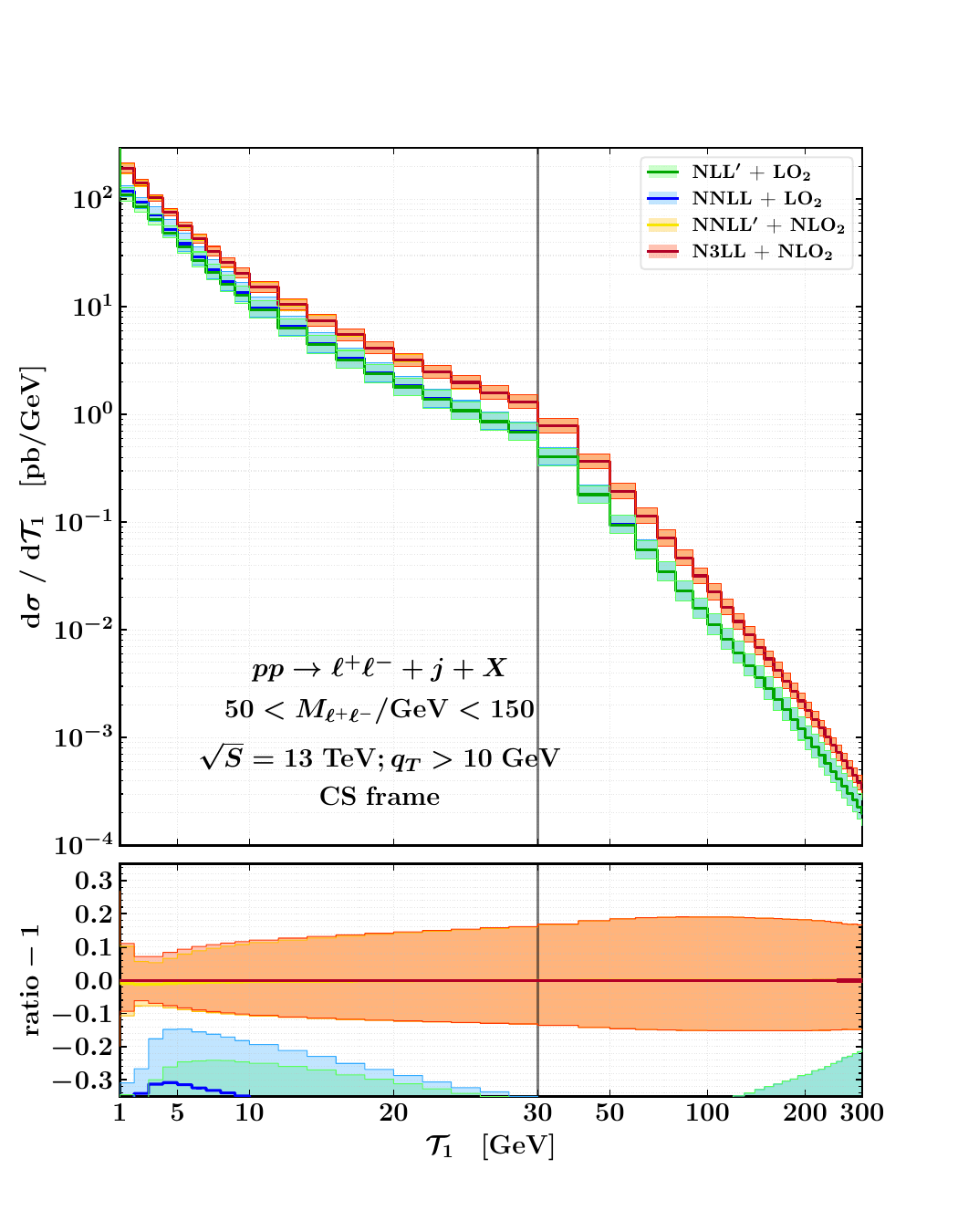}%
	\end{subfigure}
	\vspace{\spaceabovefigurecaption}
	\caption{Resummed results matched to the appropriate fixed-order on a semilogarithmic scale with $q_T > 50$~GeV (left) and with $q_T > 10$~GeV (right).}
	\label{fig:match_conv_qT}
\end{figure*}
We consider the process
\begin{align}
p p \to  ( \gamma^*/Z \to \ell^+ \ell^-) + {\text{jet}}+X \nonumber \, ,
\end{align}
at $\sqrt{S} = 13$ TeV and use the $\texttt{NNPDF31\_nnlo\_as\_0118}$ PDF
set~\cite{Ball:2017nwa}.

The factorisation and renormalisation scales are set equal to each other and equal to the dilepton transverse mass,
\begin{align}
\mu_R = \mu_F = \mu_\FO =  m_{T} \equiv \sqrt{ M^2_{\ell^+ \ell^-} + q_T^2}
\,,\end{align}
which we also use
as hard scale for the process, i.e. \mbox{$\mu_H = \mu_\FO$}. At this stage, we also fix $Q^2=s_{ab}$.

Here we report the numerical parameters used in the predictions, for ease of reproducibility.
We set the following non-zero mass and width parameters
\begin{align}
m_{\text{Z}} &= 91.1876 \, \text{GeV}\,,\qquad
&\Gamma_{\text{Z}} &= 2.4952\, \text{GeV}\,,\nn\\
m_{\text{W}} &= 80.379 \, \text{GeV}\,,\qquad
&\Gamma_{\text{W}} &= 2.0850\, \text{GeV}\,,\nn\\
m_{\text{t}} &= 173.1 \, \text{GeV}\,.\nn
\end{align}
%

In the plots presented in this section, we apply either a cut $\Tau_0 > 50$~GeV  or $q_T > 100$~GeV in order to have a well-defined Born cross section with a hard jet.
However, since our predictions depend on the choice of the cut that defines a finite Born cross section,
we study different variables and values to cut upon in \sec{resultsqT}.

\subsection{Resummed and matched predictions}
\label{sec:resummed}

In the upper panel of \fig{nonsing_Tau0_50} we show the absolute values of the
spectra for fixed-order, singular and nonsingular contributions with $\Tau_0 >
50$~GeV at different orders in the strong coupling. We plot on a logarithmic scale in
the  dimensionless $\tau_1 =  \Tau_1 / m_{T}$ variable, which is the argument
of the logarithms appearing in the cross section for our choice of  $\mu_H =
m_{T}$.
In the lower panel of the same figure we compare the
nonsingular contributions in the LAB and CS frames on a linear scale. 
At both orders one can see how the singular spectrum reproduces the fixed-order result at small values of $\tau_1$ and how the nonsingular spectrum has the expected suppressed behaviour in the $\tau_1 \to 0$ limit.
As anticipated, the nonsingular contribution in the CS frame is consistently smaller than that evaluated in the laboratory frame.
Due to the  smaller power corrections in the nonsingular contribution, from
now on we only focus on and present results in the colour-singlet frame
(though the formalism adopted is able to deal with any frame definition
related by a longitudinal boost). Similar results for  $q_T$ cuts are reported in~\sec{resultsqT}.

In the left panel of \fig{res_match_Tau0_50} (\fig{res_match_qT_100}) we show our
resummed predictions in the peak region of the $\Tau_1$ spectrum in the CS
frame, with a cut $\Tau_0 > 50$~GeV ($q_T > 100$~GeV).  We observe good
perturbative convergence between different orders.  Starting from
NNLL$^\prime$, the inclusion of  NNLO boundary conditions together with
NLO$\times$NLO mixed-terms in the factorisation formula results in a large impact on
the central values and in a sizeable decrease of the theoretical uncertainty
bands. We also notice that the differences between the N$^3$LL and the
NNLL$^\prime$ predictions are minor, suggesting  that, unlike at lower-orders,
the N$^3$LL evolution does not change considerably the NNLL$^\prime$ results.

In the right panel of \fig{res_match_Tau0_50} we present our final results
after additive matching to the fixed-order predictions. In order to better
highlight the effects in the resummation region ($\Tau_1 \lesssim 30$~GeV),
the plot is shown on a linear $\Tau_1$ scale up to $30$~GeV and a logarithmic
scale above. In this case, the addition of the nonsingular contributions
substantially modifies the resummed predictions, both in the fixed-order
($\Tau_1 \gtrsim 30$~GeV) but also in the resummation region.
This can be better appreciated by looking at
\fig{res_nons_Tau0_50}, which compares the values of the resummed and
nonsingular predictions at NNLL+LO$_2$ (left panel) and  at N$^3$LL+NLO$_2$
(right panel). The relative size of each contribution to the corresponding
matched predictions is shown in the lower inset.
We note that this poor convergence is also present when cutting on the vector
boson transverse momentum $q_T > 100$~GeV in the right panel of
\fig{res_match_qT_100} and the difference between orders grows larger
when the cut is reduced (see ~\sec{resultsqT}).
\\~\\~\\
However, since these are the first nontrivial corrections to the $\Tau_1$
spectrum, their large size is not completely unexpected and further motivates
their inclusion.

\subsection{Two-dimensional profile scales}
\label{sec:2d_profiles}
%
A final state with $N$ particles is subject to the kinematical constraint
\begin{equation}
\label{eq:ps_constraint}
\frac{\Tau_1(\Phi_N)}{\Tau_0(\Phi_N)} \leq \frac{N-1}{N} =
\begin{cases}
      1/2 \,, & N = 2 \\
      2/3 \,, & N = 3
\end{cases}
\end{equation}
where we explicitly specify the possible upper bounds that $\Tau_1/\Tau_0$
can have for the  NNLO calculation of colour-singlet plus one jet. Our goal in
this section is to formulate profile scales
that force the resummed prediction to satisfy the phase space constraint
in \eq{ps_constraint} and at the same time to have the appropriate scaling
at small and large $\Tau_1$, i.e.
\begin{align}
\mu_S(\Tau_1 \ll \mu_\FO) &\sim \Tau_1
\,, \nn \\
\mu_S(\Tau_1 \sim \mu_\FO) &\sim \mu_\FO
\,, \nn \\
\label{eq:soft_scaling_conditions}
\mu_S\bigl(\Tau_1/\Tau_0 \sim (N-1)/N \bigr) &\sim \mu_\FO
\,.\end{align}
Both requirements in \eqs{ps_constraint}{soft_scaling_conditions}
can be satisfied by formulating two-dimensional profile scales in $\Tau_1/\mu_\FO$
and $\Tau_1/\Tau_0$. To this end, we choose the soft profile scale to be
\begin{align}
\label{eq:soft_2d_prof_scale}
\mu_S \bigl( \Tau_1/\mu_\FO,& \Tau_1/\Tau_0 \bigr) \\ \nn &
= \mu_\FO \bigl[ \bigl(f_{\rm{run}}(\Tau_1/\mu_\FO) - 1\bigr) s^{(p,k)}(\Tau_1/\Tau_0) + 1\bigr]
\end{align}
where $f_{\rm{run}}$ is the same as that appearing in $\Tau_0$ profile scales used in previous \geneva implementations,
see e.g.~\refcite{Alioli:2019qzz}, while $s^{(p,k)}$ is a
logistic function
\begin{align}
\label{eq:logistic_func}
s^{(p, k)}(\Tau_1/\Tau_0)
= \frac{1}{1+e^{pk (\Tau_1/\Tau_0 - 1/p)}}
\,,\end{align}
that behaves like a smooth theta function and controls the transition to $\mu_\FO$
for a target $\Tau_1/\Tau_0$ value. It depends on the parameters $k$ and  $p$. The former fixes the slope
of the transition between canonical and fixed-order scaling, while the latter determines the transition point where this happens.
For our final predictions we use $p\!=\!2$ and $k\!=\!100$. In
\app{altprof} we further investigate the dependence of the resummed results on the way the resummation is switched off in the $\Tau_1/\Tau_0$ direction.

Finally, it is straightforward to get the beam and jet function profile scales
since they are tied to the corresponding soft profiles by
\begin{align}
\mu_B( \Tau_1/\mu_\FO, \Tau_1/\Tau_0 \bigr)
&= \sqrt{\mu_\FO \mu_S( \Tau_1/\mu_\FO, \Tau_1/\Tau_0 \bigr)}
\,, \\
\mu_J( \Tau_1/\mu_\FO, \Tau_1/\Tau_0 \bigr)
&= \sqrt{\mu_\FO \mu_S( \Tau_1/\mu_\FO, \Tau_1/\Tau_0 \bigr)}
\,,\end{align}
and for this process we set the hard scale to be
\begin{align}
\mu_H &= \mu_\FO = m_T \equiv \sqrt{ m^2_{\ell^+ \ell^-} + q_T^2}
        \,.\end{align}
When calculating scale variations we vary $\mu_\FO$ by a factor of two in either
direction. The soft, jet and beam scales variations are then calculated as
detailed in \refcite{Alioli:2019qzz} and summed in quadrature to the hard variations.

Having discussed the implementation of the resummed predictions, some freedom remains in how to treat the $\ord{\as^3}$ singular resummed-expanded term.
Since for  $\Tau_1 > \Tau_0/2$ only the real contribution $\ord{\as^3}$  with three particles can contribute in the fixed-order, one can decide to completely neglect both the resummed and the resummed-expanded terms above that threshold.  Alternatively, one can keep them both on, but with the 2D profile scales we have chosen the resummed predictions will naturally match the singular ones    for $\Tau_1 \gtrsim \Tau_0/2$ and the two contributions will cancel again in the matched predictions, leaving only the fixed-order real contribution of $\ord{\as^3}$. This behaviour is shown in \fig{tau1t_cuts_Tau0_50}, where we plot the NLO$_2$ fixed-order predictions for the $\Tau_1/\Tau_0$ ratio, together with the N$^3$LL resummed and singular ones. We include two copies of the resummed and singular predictions obtained with and without a hard cut  at  $\Tau_1/\Tau_0 = 1/2$ on the  $\ord{\as^3}$ singular contribution. This is immediately evident
from the fact that the singular prediction with this cut is zero above $\Tau_1/\Tau_0 = 1/2$. The corresponding resummed prediction does not have the same sharp jump because of the smoothing of the profile scale, but it still experiences a drastic reduction on a short $\Tau_1/\Tau_0$ range. We also notice that the singular prediction without the hard cut manifests a sudden jump: this is a consequence of the fact that for  $\Tau_1/\Tau_0 \leq 1/2$ both the $\ord{\as^2}$ and $\ord{\as^3}$ terms contribute, while above we only have the $\ord{\as^3}$ terms. The instability of this Sudakov shoulder region is also evident in the fixed-order predictions, showing the typical miscancellation between soft and collinear $\ord{\as^3}$ real emissions in the region  $\Tau_1/\Tau_0 > 1/2$. These are not compensated by their virtual counterparts, which are confined to the $\Tau_1/\Tau_0 \leq 1/2$ region.

For the predictions obtained in this work we have chosen to allow the singular
contribution at order  $\ord{\as^3}$ above $\Tau_1/\Tau_0 > 1/2$ up to the
true kinematic limit $\Tau_1/\Tau_0 = 2/3$. In principle this choice affects
the size of the $\ord{\as^3}$ nonsingular contribution across the whole
$\Tau_1$ spectrum. Therefore  we have carefully checked that our choice does
not produce numerically significant differences with the choice of imposing
$\Tau_1/\Tau_0 \leq 1/2$. In fact, for the plots shown in
\fig{nonsing_Tau0_50} we could only spot a very minor difference in the largest bins of the $\tau_1$ distribution.

\subsection{Effects of the inclusion of the $gg$ loop-induced channel}
\label{sec:ggchannel}

In this subsection we investigate the effect of the inclusion of the NLL
resummation of the $gg$ loop-induced channel in addition to the N$^3$LL+NLO$_2$ matched predictions.
Since the  $gg$ loop-induced channel starts to contribute at $\ord{\as^3}$ it is formally necessary to include it already when the resummation of the other channels is performed at NNLL$^{\prime}$ accuracy. However, as can be seen in \fig{matched_gg_Tau0_50}, its contribution is extremely small across the whole $\Tau_1$ spectrum, reaching a maximum deviation of around one per mille between 10 and 20~GeV.
The fact that this deviation is smaller than the numerical uncertainty associated with the Monte Carlo integration allows one to safely neglect this contribution.

\subsection{Results with different $\Tau_0$ and $q_T$ cuts}
\label{sec:resultsqT}
The resummation of one-jettiness requires the presence of a hard jet to have a well-defined Born cross section.
In order to investigate the effect of the selection of the hard jet here we discuss the behaviour of our preditions for different values of the $\Tau_0$ cut. We also present results obtained by requiring that the colour singlet has a substantial transverse momentum $q_T$, which is equivalent to requiring the presence of at least one hard jet with a large $k_T$ imbalance compared with other potential jets.
Lowering the  $\Tau_0$ cut value to 10 or 1~GeV, we observe a worsening of the
convergence of the resummed predictions. Moreover, the nonsingular
contribution increases with the lowering of the  $\Tau_0$ cut value and the
distance between the $\ord{\as^2}$ and $\ord{\as^3}$ contributions widens when
reaching the region $\Tau_0 \sim \Tau_1 \ll Q$. This behaviour can be easily
explained by considering that the factorisation formula in \eq{fact} has
been derived assuming $\Tau_1 \ll \Tau_0 \sim Q$. A thorough treatment of this region would necessitate a multi-differential resummation of $\Tau_0$ and $\Tau_1$, which is beyond the current state of the art~\cite{Procura:2014cba, Lustermans:2019plv}.
If we define the hard jet by placing a cut on the $q_T$ of the colour singlet system, we observe a similar behaviour when the cut is reduced. In \fig{nonsing_qT_50} we show the nonsingular contributions at $\ord{\as^2}$ and $\ord{\as^3}$ with a  $q_T > 50$~GeV cut.
Compared to the same plot for the  $\Tau_0 > 50$~GeV cut  in \fig{nonsing_Tau0_50} we observe a reduced difference between the size of the power corrections for $\Tau_1$ definitions in the two different frames.

Finally, in \fig{match_conv_qT} we show the resummed predictions matched to
the fixed-order in the peak region for  the additional cuts $q_T > 50$~GeV and
$q_T > 10$~GeV. We observe that the predictions are very sensitive to the cut
value, and the perturbative convergence is rapidly lost when decreasing the
cut value too much.

\section{Conclusions}
\label{sec:conclusions}
\begin{figure*}[ht!]
	\begin{minipage}[c]{0.49\textwidth}
		\vspace{20ex}
		\includegraphics[width=\textwidth]{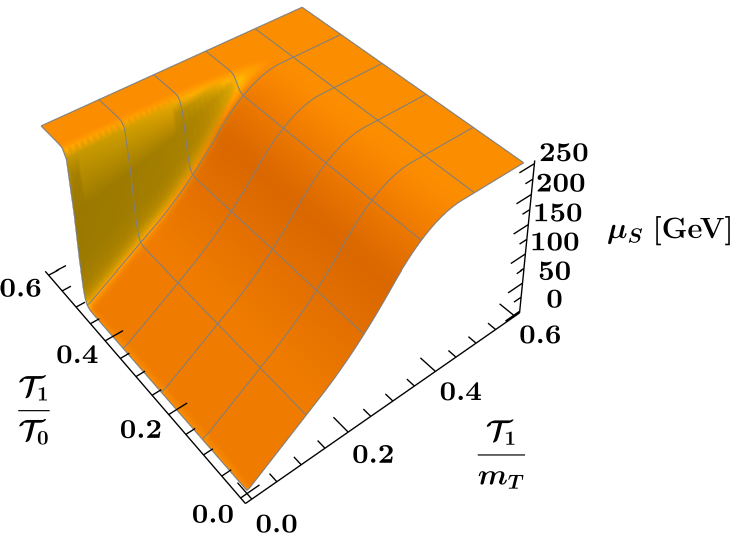}
		\vspace{2ex}
		\caption{Functional form of the two-dimensional soft profile scale.}
		\label{fig:2d_mu_soft}
	\end{minipage}
	\begin{minipage}[c]{0.49\textwidth}
		\includegraphics[width=\textwidth]{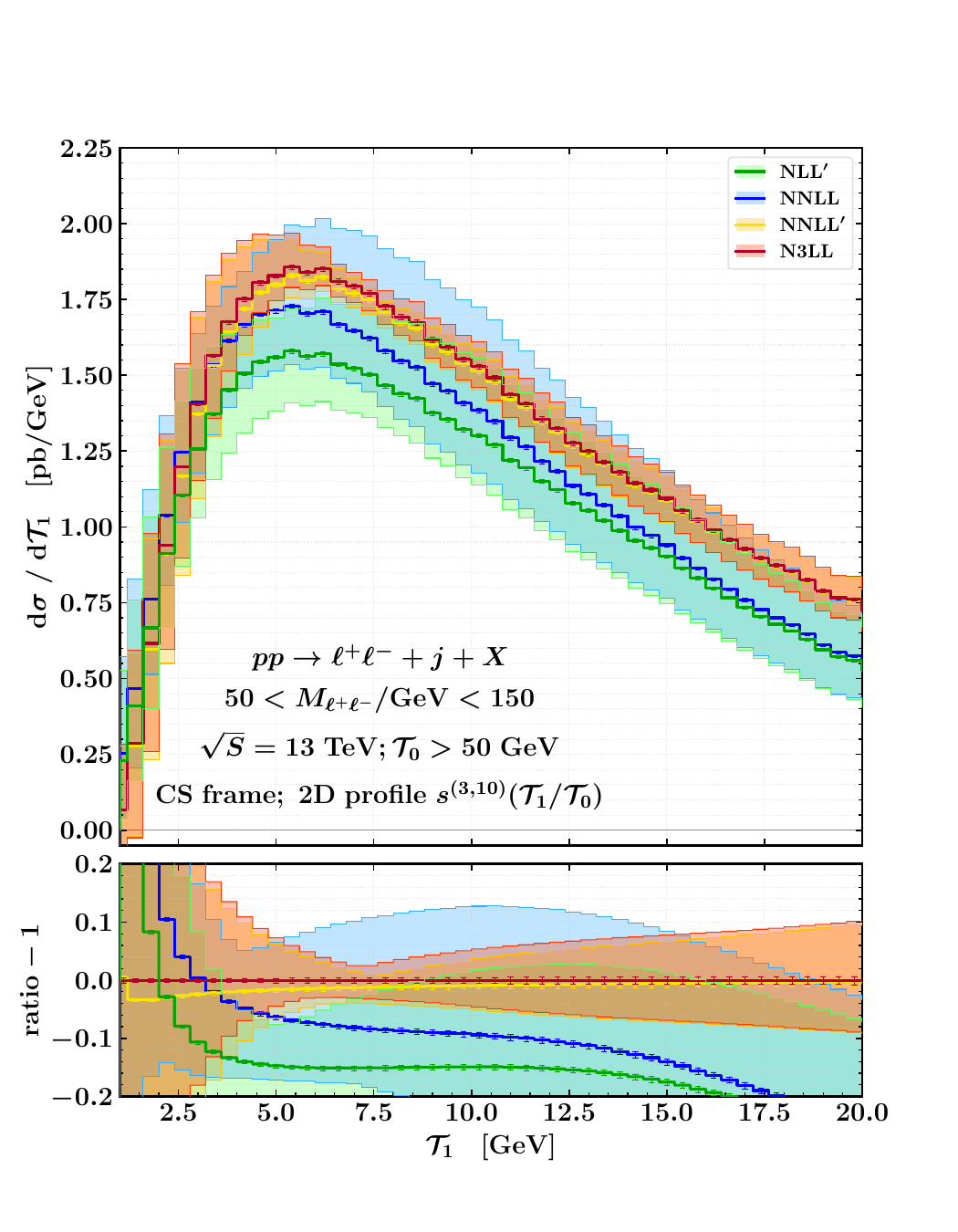}%
		\vspace{\spaceabovefigurecaption}
		\caption{Resummed results for one-jettiness distribution with $\Tau_0 > 50$~GeV at increasing accuracy, for the 2D profile with $s^{(3, 10)}(\Tau_1/\Tau_0).$}
		\label{fig:res_2D_p3k10y1_Tau0_50}
	\end{minipage}
\end{figure*}
\begin{figure*}[ht!]
	\begin{minipage}[c]{0.49\textwidth}
		\includegraphics[width=\textwidth]{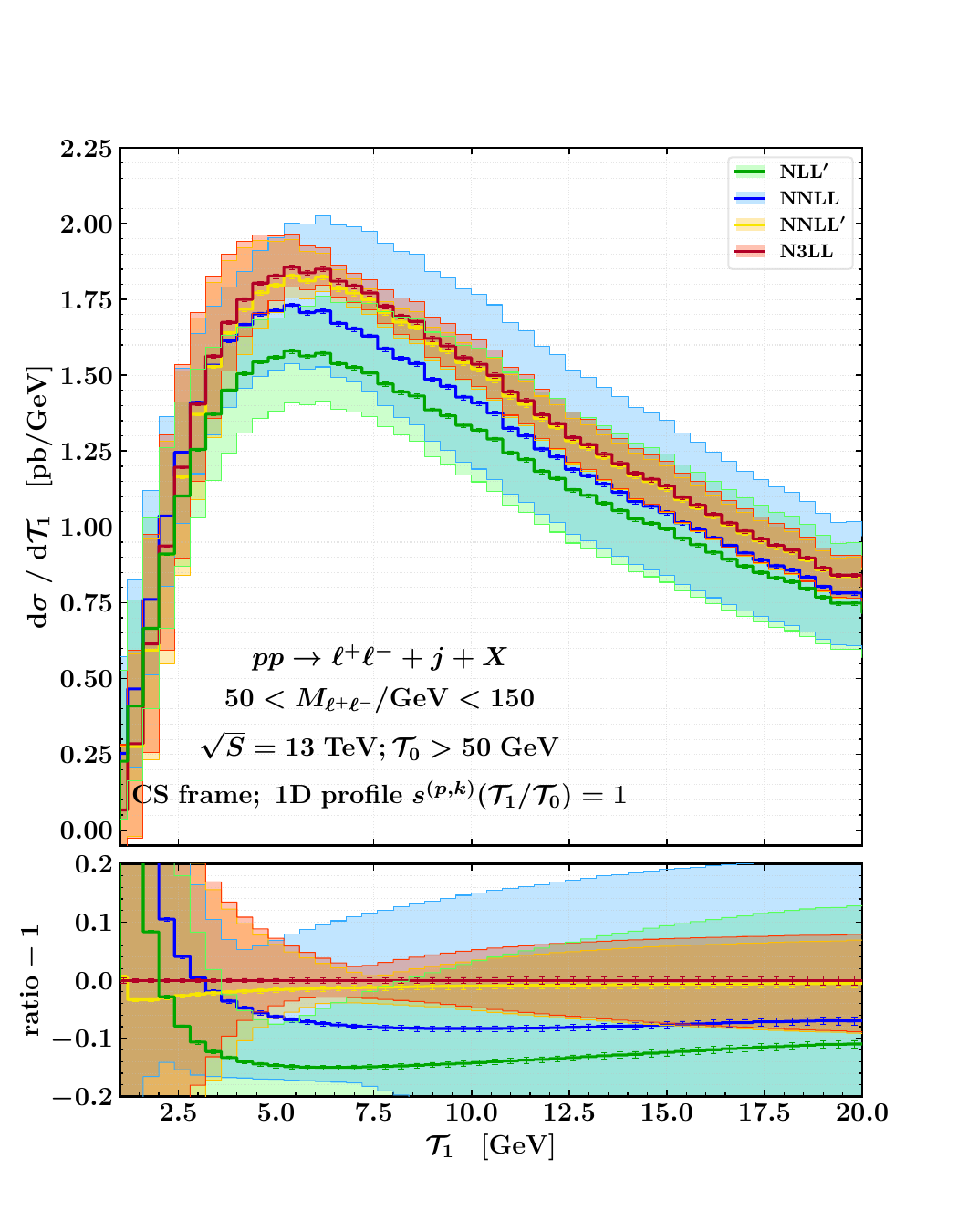}%
		\vspace{\spaceabovefigurecaption}
		\caption{Resummed results for one-jettiness distribution with $\Tau_0 > 50$~GeV at increasing accuracy, for the 1D flat profile $s^{(p, k)}(\Tau_1/\Tau_0) = 1$.}
		\label{fig:res_1D_p0_Tau0_50}
	\end{minipage}
	\begin{minipage}[c]{0.49\textwidth}
		\includegraphics[width=\textwidth]{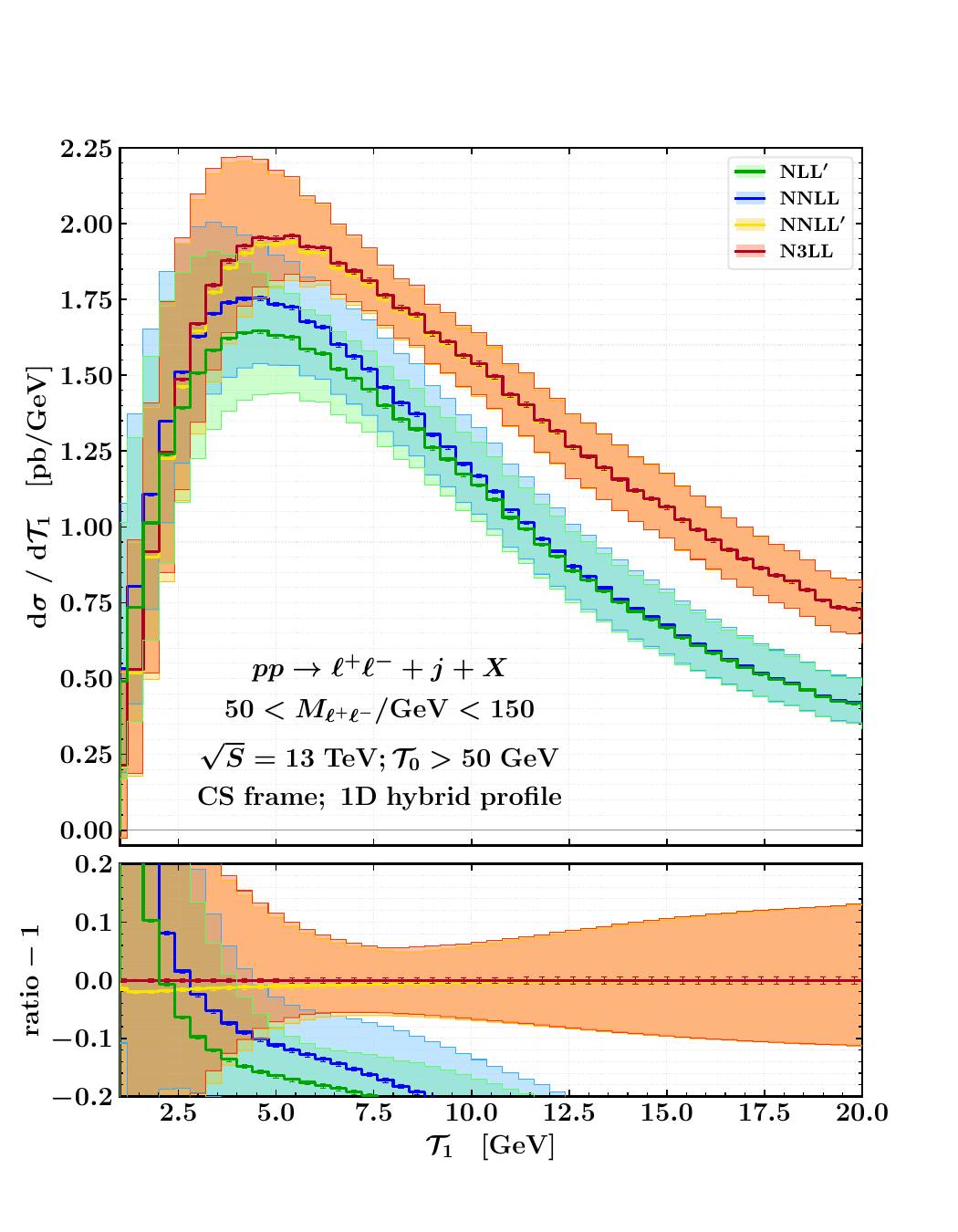}%
		\vspace{\spaceabovefigurecaption}
		\caption{Resummed results for one-jettiness distribution with $\Tau_0 > 50$~GeV at increasing accuracy, for the 1D hybrid profile discussed in the text.}
		\label{fig:res_1D_hyb_Tau0_50}
	\end{minipage}
\end{figure*}
In this work, we presented novel predictions for the $\Tau_1$ spectrum of the process \mbox{$p p \to  ( \gamma^*/Z \to \ell^+ \ell^-) + {\text{jet}}$} at NNLL$^\prime$ and N$^3$LL accuracy in resummed perturbation theory. By matching these results to the appropriate fixed-order calculation, we obtained an accurate description of the spectrum across the entire kinematic range. This is the first time that results at this accuracy have been presented for a process with three coloured partons at Born level.
Our calculation includes all two-loop singular terms as $\Tau_1\to 0$, off-shell effects of the vector bosons, the $Z/\gamma^*$ interference, as well as spin correlations.

The resummed predictions in the colour-singlet frame exhibit a good perturbative convergence, with a significant reduction of theoretical uncertainties as the perturbative order is increased. Notably, the inclusion of N$^3$LL evolution has only a minor effect on our final results.
The matching  to the fixed-order calculation was achieved by switching off the resummation  in the hard region of phase space by means of two-dimensional profile scales, which allow for the kinematic restrictions on the one-jettiness variable to be enforced consistently.
Our matched results indicate that the inclusion of the $\ord{\as^3}$ nonsingular terms is important due to their large size.

In order to assess the consistency of our implementation, we checked the explicit cancellation of the arbitrary $\mu$ dependence which appears in the separate evolution of each of the ingredients in the factorisation formula and that the singular structure of the resummed expanded results matches that of the relative fixed order. We found that, in accordance with observations in the literature, the definition of $\Tau_1$ in the laboratory frame is subject to larger nonsingular contributions. These arise due to the dependence of the observable on the longitudinal boost between the hadronic and the partonic centre-of-mass frames. To mitigate their impact, we found  that a different
definition of $\Tau_1$ (which incorporates a longitudinal boost to the frame where the vector boson has zero rapidity)
receives smaller power corrections. This makes it  suitable for slicing calculations at NNLO and for use in Monte Carlo event generators which match fixed order predictions to parton shower programs.

The $N$-jettiness variable is particularly useful in the context of constructing higher-order event generators, since it is able to act as a resolution variable which divides the phase space into exclusive jet bins. In this context, the NNLL$'$ resummed zero-jettiness spectrum has enabled the construction of NNLO+PS generators for colour-singlet production using the \geneva method. The availability of an equally accurate prediction for $\Tau_1$ in hadronic collisions will now enable these generators to be extended to cover the case of colour singlet production in association with a jet. The predictions presented in this work will be made public in a future release of \geneva.
\\~\\
\noindent {\bf Acknowledgements.}
We are grateful to our \geneva collaborators (A.~Gavardi and S.~Kallweit) for their work on the code.
We would also like to thank T.~Becher for discussions about the colour
structures of the hard anomalous dimension and F.~Tackmann for carefully reading the manuscript and for useful comments and suggestions in the presentation of the results.

The work of SA, AB, RN  and DN
has been partially supported by the ERC Starting Grant REINVENT-714788. The
work of SA and GBi is also supported by MIUR through the FARE grant R18ZRBEAFC. MAL is supported
by the UKRI guarantee scheme for the Marie Sk\l{}odowska-Curie
postdoctoral fellowship, grant ref. EP/X021416/1.
The work of D.N. is supported by the European Union - Next Generation EU (H45E22001270006).
The work of GBe was supported by the Deutsche Forschungsgemeinschaft (DFG, German Research Foundation) under grant 396021762 - TRR 257. RR is
supported by the Royal Society through Grant URF\textbackslash R1\textbackslash 201500.
GM and BD have received funding from the European Research Council (ERC)
under the European Union's Horizon 2020 research and innovation programme
(Grant agreement No. 101002090 COLORFREE). GM also acknowledges support from the Deutsche
Forschungsgemeinschaft (DFG) under Germany's Excellence Strategy -- EXC 2121
``Quantum Universe''-- 390833306 and BD from the Helmholtz Association Grant
W2/W3-116.

We acknowledge financial support, supercomputing resources and support from ICSC – Centro
Nazionale di Ricerca in High Performance Computing, Big Data and Quantum Computing –
and hosting entity, funded by European Union – NextGenerationEU.

\appendix
\section{Alternative profile scale choices}
\label{app:altprof}
In this appendix we study the dependence of the resummed results on the profile scale definition in the $(\Tau_1/m_T, \Tau_1/\Tau_0)$ plane. We start by showing the functional form of $\mu_S$ for the default 2D profile scales in  \fig{2d_mu_soft}. As observed, when the LO$_2$ kinematical constraint~\eq{ps_constraint} is satisfied, the factor  $s^{(2, 100)}(\Tau_1/\Tau_0 \lesssim 1/2) \to 1$ in \eq{soft_2d_prof_scale} and therefore the scaling of
$\mu_S$ is entirely dictated by $f_{\rm{run}}$, depending only on the value of
$\Tau_1/\mu_\FO$.
On the other hand, when the $\Tau_1/\Tau_0 \leq 1/2 $ condition is violated, $s^{(2, 100)}(\Tau_1/\Tau_0 \gtrsim 1/2) \to 0$,
which implies that $\mu_S = \mu_\FO$.
This is a crucial asymptotic limit,
since for \mbox{$\Tau_1/\Tau_0 \gtrsim 1/2$} the fixed-order and singular cross sections pass a kinematical boundary. Therefore, since the  fixed-order corrections are extremely relevant in that region, care must be taken to switch off the $\Tau_1$ resummation before passing the same threshold.

In \fig{res_2D_p3k10y1_Tau0_50} we show resummed predictions obtained using 2D
profiles with $p\!=\!3$ and $k\!=\!10$, which results in a earlier and
smoother switch-off of the resummation around  $\Tau_1/\Tau_0\sim 1/3$. As one
can see by comparing the results with the left panel of
\fig{res_match_Tau0_50}, by doing so the convergence of successive
perturbative orders is slightly worsened.
Alternatively, we have explored the usage of 1D profile scales, either by
removing the suppression in the  $\Tau_1/\Tau_0$ direction, see
\fig{res_1D_p0_Tau0_50}, or by switching off the resummation
in the   $\Tau_1/\Tau_0$ direction by means of a 1D hybrid profile, see
\fig{res_1D_hyb_Tau0_50}. The hybrid profile approach has previously
been successfully used in enforcing multi-differential profile scales switch-offs~\cite{Cal:2023mib}.
In our case it is defined by
\begin{align}
	\label{eq:soft_1d_hyb_prof_scale}
	\mu_S\bigl( \Tau_1/\mu_\FO,& \Tau_1/\Tau_0 \bigr) \\\nn&
	=  \mu_\FO f_{\rm{run}}(\Tau_1/\Tau_0) + \Tau_1 \bigl( 1 - f_{\rm{run}}(\Tau_1/\Tau_0)\bigr)\,,
\end{align}
where now the argument of  $f_{\rm{run}}$ is the ratio $\Tau_1/\Tau_0$ rather than $\Tau_1/\mu_\FO$. The formula in \eq{soft_1d_hyb_prof_scale} smoothly interpolates between $\Tau_1$ and $\mu_\FO$ on a diagonal slice of the $(\Tau_1/\Tau_0, \Tau_1/m_T)$ plane. 
In \fig{res_1D_p0_Tau0_50} we observe that removing the $\Tau_1/\Tau_0$ suppression has very small effects on the $\Tau_1$ distribution, maintaining a good perturbative convergence across orders. This, however, does not provide the correct suppression of the resummation effects past the kinematic endpoint in the  $\Tau_1/\Tau_0$ direction.
The usage of the hybrid profile shows instead a much poorer convergence (see \fig{res_1D_hyb_Tau0_50}).
In particular, we notice a change in the resummed predictions also in the peak region, which should follow a canonical scaling.
This is easily explained by the fact that for the hybrid profiles in \eq{soft_1d_hyb_prof_scale} the $\mu_S$ behaviour at low $\Tau_1$ is changed from $\Tau_1$ to $(1+m_T/\Tau_0) \Tau_1$, which is still a canonical scaling but includes small artificial leftover logarithms.

\section{Resummed formula at N$^3$LL accuracy}
\label{app:resn3ll}
  In this section we report the full formula for the  N$^3$LL resummation with
  the explicit combination of the hard, soft, beam and jet boundary terms,
  evaluated at the appropriate  order, for completeness. It reads 
\begin{widetext}
  \begin{align}
    \label{eq:n3llformula-long}
    \frac{\de\sigma^{\textsc{n}^3\textsc{ll}}}{\de\Phi_1\de\Tau_1}  = & \sum_{\kappa} 
\exp\bigg\{
4(C_{a}+C_{b})
  K^{\textsc{n}^3\textsc{ll}}_{\Gamma_{\cusp}}(\mu_B, \mu_H) + 4 C_{c}
  K^{\textsc{n}^3\textsc{ll}}_{\Gamma_{\cusp}}(\mu_J, \mu_H) 
  - 2 (C_{a}+C_{b}
  + C_{c})
  K^{\textsc{n}^3\textsc{ll}}_{\Gamma_{\cusp}}(\mu_S, \mu_H) \nn \\&\quad
  \qquad \quad - 2 C_{c} L_J\ \eta^{\textsc{n}^3\textsc{ll}}_{\Gamma_{\cusp}}(\mu_J, \mu_H)
  -2 ( C_{a} L_B 
  + C_{b} L'_B ) \eta^{\textsc{n}^3\textsc{ll}}_{\Gamma_{\cusp}}(\mu_B, \mu_H)
   + K^{\textsc{n}^3\textsc{ll}}_{\gamma_{\rm tot}} \nn \\&\quad
  + \bigg[ 
  C_{a} \ln\left(\frac{Q^2_a u }{s t} \right)
  + C_{b} \ln\left( \frac{Q^2_b t }{s u }\right) 
  + C_{\kappa_j} \ln\left( \frac{Q^2_J s }{t u }\right)  
  +  (C_{a}+C_{b}+C_{c}) L_S \bigg] \eta^{\textsc{n}^3\textsc{ll}}_{\Gamma_{\cusp}}(\mu_S, \mu_H) 
  \nn \\
& +  \sum_{R=F,A}\bigg[ 8\, \big(D_{aR}+D_{bR}\big) K^{\textsc{n}^3\textsc{ll}}_{g^R}(\mu_B,\mu_H) + 8\,  D_{cR} K^{\textsc{n}^3\textsc{ll}}_{g^R}(\mu_J,\mu_H) \, \nn \\
&\qquad - 4\,  \big(D_{aR}+D_{bR}+D_{cR}\big) K^{\textsc{n}^3\textsc{ll}}_{g^R}(\mu_S,\mu_H) \,  - 4 D_{cR} L_J  \eta^{\textsc{n}^3\textsc{ll}}_{g^R}(\mu_J,\mu_H)\,  - 4\, \big(D_{aR} L_B + D_{bR} L^\prime_B\big) \eta^{\textsc{n}^3\textsc{ll}}_{g^R}(\mu_B,\mu_H) \nn \\
& \qquad + 2\, \bigg[D_{aR} \ln \bigg(\frac{Q^2_a u}{s t}\bigg)+D_{bR} \ln
  \bigg(\frac{Q^2_b t}{s u}\bigg) + D_{cR} \ln \bigg(\frac{Q^2_J s}{t
  u}\bigg)\, + \big(D_{aR}+D_{bR}+D_{cR}\big)
  L_S\bigg]\eta^{\textsc{n}^3\textsc{ll}}_{g^R}(\mu_S,\mu_H)\bigg] \bigg\}
\nn\\
&\times \bigg\{   H^{(0)}_{\kappa}(\Phi_1, \mu_H) \bigg[ f_{\kappa_a}(x_a,\mu_B) f_{\kappa_b}(x_b,\mu_B) \bigg( 1 +  \frac{\as(\mu_S)}{4 \pi} 
  \tilde{S}^{\kappa\, (1)} \left(\partial_{\eta_S}+ L_S, \mu_S\right)  
   \nn \\ 
  & \qquad +  \frac{\as(\mu_J)}{4 \pi}  \tilde{J}^{(1)}_{\kappa_J}(\partial_{\eta_J}+L_J,\mu_J) + \frac{\as(\mu_S) \as(\mu_J)}{16 \pi^2}    \tilde{J}^{(1)}_{\kappa_J}(\partial_{\eta_J}+L_J,\mu_J) \tilde{S}^{\kappa\, (1)} \left(\partial_{\eta_S}+ L_S, \mu_S\right)  \nn \\ 
  & \qquad + \frac{\as^2(\mu_S)} {16 \pi^2}  \tilde{S}^{\kappa\, (2)} \left(\partial_{\eta_S}+ L_S, \mu_S\right) + \frac{\as^2(\mu_J)}{16 \pi^2} 
 \tilde{J}^{(2)}_{\kappa_J}(\partial_{\eta_J}+L_J,\mu_J) \bigg)
    \nn \\
  & \qquad  + \bigg( \frac{\as(\mu_B)}{4 \pi}  \tilde{B}^{(1)}_{\kappa_a}(\partial_{\eta_B} + L_B,x_a,\mu_B) \Big( 1 + \frac{\as(\mu_S)}{4 \pi} 
 \tilde{S}^{\kappa\, (1)} \left(\partial_{\eta_S}+ L_S, \mu_S\right)   \nn \\ & \qquad +   \frac{\as(\mu_J)}{4 \pi}  \tilde{J}^{(1)}_{\kappa_J}(\partial_{\eta_J}+L_J,\mu_J) \Big) + \frac{\as^2(\mu_B)}{16 \pi^2} \tilde{B}^{(2)}_{\kappa_a}(\partial_{\eta_B} + L_B,x_a,\mu_B) \bigg)
f_{\kappa_b}(x_b,\mu_B)   
  \nn \\ &
 \qquad  + f_{\kappa_a}(x_a,\mu_B) \bigg( \frac{\as^2(\mu_B)}{16 \pi^2} \tilde{B}^{(2)}_{\kappa_b}(\partial_{\eta_B'}+L_B',x_b,\mu_B)  + \frac{\as(\mu_B)}{4 \pi} \tilde{B}^{(1)}_{\kappa_b}(\partial_{\eta_B'}+L_B',x_b,\mu_B) \times \nn \\ &
   \qquad \quad \Big( 1 +  \frac{\as(\mu_S)}{4 \pi} \tilde{S}^{\kappa\, (1)} \left(\partial_{\eta_S}+ L_S, \mu_S\right) + \frac{\as(\mu_J)}{4 \pi} \tilde{J}^{(1)}_{\kappa_J}(\partial_{\eta_J}+L_J,\mu_J) \Big) \bigg) \bigg]
   \nn \\ &
   \quad  + \frac{\as(\mu_H)}{4 \pi} H^{(1)}_{\kappa}(\Phi_1, \mu_H) \bigg[ f_{\kappa_a}(x_a,\mu_B) f_{\kappa_b}(x_b,\mu_B) \bigg( 1 + \frac{\as(\mu_S)}{4 \pi}
  \tilde{S}^{\kappa\, (1)} \left(\partial_{\eta_S}+ L_S, \mu_S\right)
    \nn \\
  & \qquad  + \frac{\as(\mu_J)}{4 \pi} \tilde{J}^{(1)}_{\kappa_J}(\partial_{\eta_J}+L_J,\mu_J)  \bigg) + \bigg( \frac{\as(\mu_B)}{4 \pi} \tilde{B}^{(1)}_{\kappa_a}(\partial_{\eta_B} + L_B,x_a,\mu_B) 
f_{\kappa_b}(x_b,\mu_B)    
   \nn \\ &
   \quad  + \frac{\as(\mu_B)}{4 \pi} f_{\kappa_a}(x_a,\mu_B)  
\tilde{B}^{(1)}_{\kappa_b}(\partial_{\eta_B'}+L_B',x_b,\mu_B) \bigg) \bigg] 
\nn\\& \quad
+ \frac{\as^2(\mu_H)}{16 \pi^2} H^{(2)}_{\kappa}(\Phi_1, \mu_H)  f_{\kappa_a}(x_a,\mu_B) f_{\kappa_b}(x_b,\mu_B)\bigg\}
 \times
\frac{Q^{-\eta^{{\textsc{n}^3\textsc{ll}}}_{\rm tot}}}{{\Tau_1}^{1-\eta^{{\textsc{n}^3\textsc{ll}}}_{\rm tot}}}\, \frac{\eta^{{\textsc{n}^3\textsc{ll}}}_{\rm tot}\ e^{-\gamma_E \eta^{{\textsc{n}^3\textsc{ll}}}_{\rm tot}}}{\Gamma(1+\eta^{{\textsc{n}^3\textsc{ll}}}_{\rm tot})} \,.
\end{align}
\end{widetext}

\bibliographystyle{apsrev4-2}
\bibliography{geneva}
\end{document}